\documentclass[12pt]{article}
\usepackage{ulem}
\usepackage{amsfonts}
\usepackage{amsmath,amsthm,amssymb}
\usepackage{mathptmx}
\usepackage{geometry}
\usepackage{graphicx}
\usepackage{subfigure}
\usepackage{hyperref}
\usepackage{cancel}
\usepackage{textcomp}
\usepackage{tikz}
\usepackage{bm}
\usepackage{mathrsfs}
\usepackage{filecontents}
\usepackage{times}
\usepackage{epsfig}
\usepackage{xcolor}
\usepackage{slashed}
\usepackage{booktabs}
\usepackage{latexsym}
\usepackage{verbatim}
\usepackage{extarrows}
\usepackage{multirow}
\usepackage{rotating}
\usepackage{colortbl}
\usepackage{indentfirst}
\usepackage[numbers,sort&compress]{natbib}
\usepackage{soul}
\definecolor{mygray}{gray}{.9}
\definecolor{intnull}{RGB}{213,229,255}
\usepackage{arydshln}
\usepackage{diagbox}
\usepackage{appendix}

\newcommand{\dif}{\mathrm{d}}

\newcommand{\Hor}{\mathrm{H}}
\newcommand{\Min}{\mathrm{min}}

\usepackage{caption}
\usepackage{tikz}
\usetikzlibrary{arrows,shapes,chains}
\begin{document}
\renewcommand{\thefootnote}{\fnsymbol{footnote}}
\baselineskip=16pt
\pagenumbering{arabic}
\vspace{1.0cm}
\begin{center}
{\Large\sf Preliminary analyses on dynamics and thermodynamics of rotating regular black holes}
\\[10pt]
\vspace{.5 cm}
{Hao Yang\footnote{E-mail address: hyang@mail.nankai.edu.cn}}, 
{Chang-Jiang Yu\footnote{E-mail address: 2120210161@mail.nankai.edu.cn}}, and
{Yan-Gang Miao\footnote{Corresponding author. E-mail address: miaoyg@nankai.edu.cn}}
\vspace{3mm}

{School of Physics, Nankai University, Tianjin 300071, China}

\vspace{4.0ex}
\end{center}
\begin{center}
{\bf Abstract}
\end{center}

We investigate the dynamic and thermodynamic laws governing rotating regular black holes. 
By analyzing dynamic properties, i.e., the interaction between scalar particles and rotating regular black holes, we establish the criteria that determine whether such black holes satisfy the laws of thermodynamics or not. 
In addition, we provide the general form of conserved quantities related to rotating regular black holes, including the relevant flows associated with neutral scalar particles.
Meanwhile, we reexamine the relationship between the third law of thermodynamics and  weak cosmic censorship conjecture for rotating regular black holes.
In accordance with the criteria mentioned above, we discuss the laws of thermodynamics for three models of rotating regular black holes: Rotating Hayward black holes, Kerr black-bounce solutions, and loop quantum gravity black holes. 
Our findings indicate that none of the three models satisfies the first law of thermodynamics. 
In particular, the first and third models fail to comply with the three laws of thermodynamics, while the second model satisfies only the second and third laws of thermodynamics.
Finally, we attempt to rescue the laws of thermodynamics by modifying entropy or extending phase space. 
However, the two scenarios are not able to ensure the three laws of thermodynamics in the three models, which reveals an unusual property of rotating regular black holes.

\renewcommand{\thefootnote}{\arabic{footnote}}
\newpage
\tableofcontents

\newpage
\section{Introduction}
\label{sec:intr}
As the most successful gravitational theory, general relativity has been confirmed~\cite{LIGOScientific:2016aoc,EventHorizonTelescope:2019dse,EventHorizonTelescope:2019ths} by various recent astronomical observations.
Currently, the primary challenge that general relativity confronts remains the issue of spacetime singularity.
The singularity theorem proposed~\cite{Hawking:1970zqf,Penrose:1964wq,Hawking:1973uf} by Hawking and Penrose points out that singularities exist inevitably in the spacetime under certain conditions.
The presence of singularities disrupts the coherence and self-consistency of spacetime, but singularities are always hidden by event horizons and remain unobservable to external observers. This is known as the weak cosmic censorship conjecture~\cite{Penrose:1969pc}.
As far back as the previous century, gedanken experiments were employed~\cite{WALD1974548} to assess the rationality of the weak cosmic censorship conjecture.
The primary objective of gedanken experiments is to examine~\cite{Jacobson:2009kt,Sorce:2017dst,Hubeny:1998ga,Hod:2002pm} the possibility of event horizon's destruction by particle injection, giving the result~\cite{Barausse:2010ka,Gwak:2018akg,Gwak:2021tcl,Sorce:2017dst} that the event horizons of Kerr and Kerr-Newman black holes with singularities
are not destroyed.

The weak cosmic censorship conjecture implies that an observational boundary of black holes  is confined on black hole's event horizons.
Therefore, the mechanical properties of black holes are commonly described through various quantities that take values on an event horizon, including but not limited to area, surface gravity, and angular velocity.
In particular, the mechanical laws governing black holes are derived~\cite{Bardeen:1973gs} from interrelationships among these quantities. Moreover,
the Hawking temperature establishes the correspondence between mechanical and thermodynamic quantities of black holes, leading to the construction of thermodynamics laws for black holes.
As a particle incidence to a black hole can induce changes in mechanical quantities, such as area, mass, and angular momentum, one can employ the  gedanken experiments mentioned above to verify~\cite{Gwak:2018akg,Gwak:2021tcl} the thermodynamic laws of black holes. As the thermodynamic quantities, such as entropy and temperature, are defined solely on horizons,  
the black hole thermodynamics cannot be established if horizons are absent.
In other words, the weak cosmic censorship conjecture ensures the establishment of thermodynamic laws for black holes.
Conversely, the thermodynamic laws substantiate the rationality of the weak cosmic censorship conjecture.
Therefore, the cosmic censorship conjecture and thermodynamic laws of black holes are complementary to each other.

Recently, diverse techniques have been employed~ \cite{Balart:2014cga,Ayon-Beato:1998hmi,Hayward:2005gi,Nicolini:2005vd,Simpson:2018tsi,Bodendorfer:2019jay,Bodendorfer:2019nvy,Dymnikova:1992ux,Bronnikov:2005gm,Bronnikov:2006fu,Ayon-Beato:2000mjt,Franzin:2021vnj,Simpson:2019mud,Bonanno:2000ep,Koch:2014cqa,Bouhmadi-Lopez:2020wve} to construct static and spherically symmetric regular black holes that have no essential singularities,
and can be summarized into three categories: 
\begin{itemize}
    \item  To solve~\cite{Dymnikova:1992ux,Hayward:2005gi,Bronnikov:2005gm,Bronnikov:2006fu,Nicolini:2005vd} the Einstein field equations under a special symmetry or matter source.
    Among the black holes constructed in this category, the Hayward black hole is a typical example~\cite{Hayward:2005gi} where the vacuum energy density distribution is considered in Einstein's gravity;
    
    \item To  modify~\cite{Simpson:2018tsi,Franzin:2021vnj,Simpson:2019mud,Ayon-Beato:1998hmi,Ayon-Beato:2000mjt,Ayon-Beato:2004ywd,Balart:2014cga} a metric directly so that the corresponding spacetime has no singularity, and then to deduce a possible matter source inversely.
    The key step is to find out a novel metric function that removes singularities in spacetime, which was previously achieved by experience and formulated by specific functions.
    Recently, a systematic modification of metrics has been proposed~\cite{Simpson:2018tsi,Franzin:2021vnj} in order to construct a regular black hole.
    Applying such a modification to the Schwarzschild metric, one can derive the so-called black-bounce metric, and verify~\cite{Bronnikov:2021uta} that this metric is the solution of Einstein's gravity coupled with the phantom scalar field and the electromagnetic field, where the matter source is deduced inversely from the metric;
    
    \item To solve~\cite{Bonanno:2000ep,Koch:2014cqa,Bouhmadi-Lopez:2020wve,Bodendorfer:2019jay,Bodendorfer:2019nvy,Bojowald:2020dkb} a metric for regular black holes in the framework of modified theories of gravity. 
    Among them, the loop quantum gravity~\cite{Bojowald:2020dkb} is a nonperturbative theory and goes beyond general relativity, and it can resolve  singularities in cosmological and black hole spacetimes.
    Therefore, regular black holes can be constructed~\cite{Bonanno:2000ep,Bonanno:2020fgp} naturally under this theory.
\end{itemize}
As an astronomical black hole is rotating in nature, the Newman-Janis algorithm (NJA) has been applied and improved to extend~\cite{Smailagic:2010nv,Modesto:2010rv,Modesto:2010uh,Caravelli:2010ff,Bambi:2013ufa,Azreg-Ainou:2014aqa,Azreg-Ainou:2014pra} the investigation of regular black holes from the static and spherically symmetric case to the rotating and axially symmetric case.
Owing to the absence of singularities, the primary significance of the weak cosmic censorship conjecture does not lie in avoiding naked singularities, but rather in upholding  thermodynamic laws.
However, a recent paper on rotating loop quantum gravity black holes has claimed~\cite{Yang:2022yvq} that the incidence of scalar particles into the near-extreme configuration of rotating loop quantum gravity black holes can disrupt event horizons.
Although it is a special case, this result implies~\cite{Yang:2022yvq} that the thermodynamics of regular black holes may differ from that of singular black holes.
In fact, only metric singularities were avoided in the construction of regular black holes in early literature, while the compatibility with thermodynamic laws was neglected.
Therefore, it remains unsolved whether regular black holes adhere to the thermodynamic laws that are valid to singular black holes.

In the present work  we focus on investigating the behavior of rotating regular black holes coupled with scalar particles, and propose the criteria for establishing the relevant thermodynamic laws, where these criteria match the self-consistency in constructing rotating regular black holes.
Moreover, we examine the correlation between the third law of thermodynamics and weak cosmic censorship conjecture, yielding different conclusions from those  in Kerr and Kerr-Newman black holes.
Finally, we analyze whether rotating regular black holes satisfy the  thermodynamic laws deduced from singular black holes through illustrative examples and try to rescue these laws by modifying entropy or extending phase spaces.

The outline of this paper is as follows.
In Sec.~\ref{sec:Rotating}, we provide a brief introduction to rotating regular black holes constructed by the revised NJA, and present the mass and angular momentum in their general forms.
In Sec.~\ref{sec:scalar}, we discuss the behavior of a scalar field near a rotating regular black hole and derive the scalar field flux and the Hawking temperature of rotating regular black holes.
In Sec.~\ref{sec:thermo}, we establish the criteria to give the laws of thermodynamics for rotating regular black holes, and discuss the relationship between the third law of thermodynamics and weak cosmic censorship conjecture.
In Sec.~\ref{sec:sp_model}, we apply these criteria to three models of rotating regular black holes and explore the possibility of their fulfillment by modifying entropy or extending phase spaces.
Finally, we give our summary and outlook in Sec.~\ref{sec:conclusion}.

\section{Rotating regular black holes}
\label{sec:Rotating}
\subsection{General form of metrics}
The construction of rotating regular black holes involves~\cite{Modesto:2010rv,Brahma:2020eos,Mazza:2021rgq,Abdujabbarov:2016hnw} the use of the NJA~\cite{Newman:1965,Gurses:1975,Drake:1998gf,Azreg-Ainou:2014aqa,Azreg-Ainou:2014pra}, an algebraic method for transforming a static and spherically symmetric black hole solution into a rotating and  axially symmetric one. 
Now we provide a brief introduction to this method.
Initially, we consider a general static and spherically symmetric metric,
\begin{equation}
ds_{\rm static}^2=-F(r)dt^2+\frac{dr^2}{F(r)}+H(r)\left(d\theta^2+\sin^2\theta d\varphi^2\right).\label{lineelemstatic}
\end{equation}
Because the spacetime under consideration is asymptotically flat at infinity, the functions $F(r)$ and $H(r)$ should satisfy the following conditions,
\begin{equation}\label{eq:lim FH}
\lim_{r\to\infty}F(r)=1,\qquad \lim_{r\to\infty}H(r)=r^2.
\end{equation}
Therefore, we expand $F(r)$ and $H(r)$ as series of $1/r$ near infinity, $r\rightarrow\infty$, as follows:
\begin{equation}\label{eq:series F}
\lim_{r\rightarrow\infty}F(r)=\sum_{n=0}^{\infty}a_nr^{-n},\qquad a_n=\frac{1}{n!}\frac{\partial^n}{\partial(1/r)^n}F(r)\Big|_{r\to\infty},\qquad a_0=1;
\end{equation}
\begin{equation}\label{eq:series H}
\lim_{r\rightarrow\infty}H(r)=\sum_{n=0}^{\infty}b_nr^{2-n},\qquad b_n=\frac{1}{n!}\frac{\partial^n}{\partial(1/r)^n}\frac{H}{r^2}\Big|_{r\to\infty},\qquad b_0=1 .
\end{equation}
In the advanced null coordinates, $(u, r, \theta, \phi)$, defined by
\begin{equation}
\dif u=\dif t-\frac{\dif r}{F(r)},
\end{equation}
the contravariant form of the metric can be expressed in terms of a null tetrad as
\begin{equation}
g^{\mu\nu}=-l^\mu n^\nu-l^\nu n^\mu+m^\mu m^{*\nu}+m^\nu m^{*\mu},
\end{equation}
where 
\begin{subequations}\label{eq:tetrad}
\begin{equation}
l^\mu=\delta^\mu_r,
\end{equation}
\begin{equation}
n^\mu=\delta^\mu_u-\frac{F}{2}\delta^\mu_r,
\end{equation}
\begin{equation}
m^\mu=\frac{1}{\sqrt{2H(r)}}\left(\delta^\mu_\theta+\frac{i}{\sin\theta}\delta^\mu_\phi\right),
\end{equation}
\begin{equation}
l_\mu l^\mu=m_\mu m^\mu=n^\nu n_\nu=l_\mu m^\mu=n_\mu m^\mu=0,
\end{equation}
\begin{equation}
l_\mu n^\mu=-m_\mu m^{*\mu}=1,
\end{equation}
\end{subequations}
and ``$*$" denotes complex conjugate.
The rotation is introduced via the complex transformation,
\begin{equation}
r\rightarrow r+ia\cos\theta,\qquad  u\rightarrow u-i a\cos\theta,
\end{equation}
where $a$ is rotation parameter, and $\delta^\mu_\nu$ is required to transform as a vector under the above complex transformation,
\begin{equation}\label{eq:transform-delta}
\delta^\mu_r\rightarrow \delta^\mu_r,
\qquad \delta^\mu_u\rightarrow\delta^\mu_u,
\qquad \delta^\mu_\theta\rightarrow\delta^\mu_\theta+ia\sin\theta(\delta^\mu_u-\delta^\mu_r),
\qquad \delta^\mu_\phi\rightarrow\delta^\mu_\phi.
\end{equation}
After the above fulfillment, $\{F, H\}$ are generalized to $\{B, \Psi\}$ with rotation:
\begin{equation}\label{eq:transform-function}
\{F(r),H(r)\}\rightarrow\{B(r,\theta,a),\Psi(r,\theta,a)\},
\end{equation}
where $\{B, \Psi\}$ are real functions to be determined and should recover their static counterparts in the limit of $a\rightarrow 0$, i.e., 
\begin{equation}
\lim_{a\to 0}B(r,\theta,a)=F(r),\qquad \lim_{a\to 0}\Psi(r,\theta,a)=r^2.
\end{equation}
In particular, the line element without rotation, see Eq.~(\ref{lineelemstatic}), ia now transformed to the one with rotation when Eqs.~\eqref{eq:transform-delta} and ~\eqref{eq:transform-function} are considered, 
\begin{equation}
\begin{split}
\dif s^2=& -B\dif u^2-2\dif u\dif r-2a\sin^2\theta\left(1-B\right)\dif u\dif \phi+2a\sin^2\theta\dif r\dif \phi\\
& +\Psi\dif\theta^2+\sin^2\theta\left[\Psi+a^2\sin^2\theta\left(2-B\right)\right]\dif\phi^2.
\end{split}
\end{equation}

Next, we rewrite the above line element with the Boyer-Lindquist coordinates, and let the metric have only one non-vanishing off-diagonal term, $g_{t\varphi}$.
To reach the goal, we need the following coordinate transformation,
\begin{equation}
\dif u=\dif t+\lambda(r)\dif r,\qquad \dif\phi=\dif\varphi+\chi(r)\dif r,
\end{equation}
where $\{\lambda(r), \chi(r)\}$ depend only on $r$ in order to ensure integrability.
If the transformation Eq.~\eqref{eq:transform-function} is given a priori , $\{\lambda(r), \chi(r)\}$ may not exist.
Considering these constraints, we determine the formulations of $\{B(r, \theta, a), \Psi(r, \theta), \lambda(r), \chi(r)\}$,
\begin{subequations}
\begin{equation}
B(r,\theta)=\frac{FH+a^2\cos^2\theta}{\Psi},
\end{equation}
\begin{equation}
\Psi(r,\theta)=H+a^2\cos^2\theta,
\end{equation}
\begin{equation}
\lambda(r)=-\frac{H+a^2}{FH+a^2},
\end{equation}
\begin{equation}
\chi(r)=-\frac{a}{FH+a^2}.
\end{equation}
\end{subequations}
As a result, we obtain the line element for rotating regular black holes with the Kerr-like form,
\begin{equation}\label{eq:rotating_1}
ds^2=\frac{\Psi}{\Sigma}\left[-\left(1-\frac{fH}{\Sigma}\right)dt^2+\frac{\Sigma}{\Delta}dr^2-\frac{2f H a}{\Sigma}\sin^2\theta dtd\phi+\Sigma d\theta^2+ \frac{A}{\Sigma}\sin^2\theta d\phi^2\right],
\end{equation}
where 
\begin{eqnarray}
\Sigma&=&H+a^2\cos^2\theta,\\
f&=&1-F,\\
\Delta&=&FH+a^2,\label{defDelta}\\
A&=&(H+a^2)^2- a^2\Delta\sin^2\theta.
\end{eqnarray}
Although the choice of $\Psi$ has certain degrees of freedom, $\Psi=\Sigma$ is  typically selected~\cite{Simpson:2019mud,Brahma:2020eos,Modesto:2010rv,Bambi:2013ufa,Azreg-Ainou:2014pra} for a rotating regular black hole. 
In this situation, we rewrite Eq.~\eqref{eq:rotating_1} in another form,
\begin{equation}\label{rotating_metric}
ds^2=-\frac{\Delta \Sigma}{A}dt^2+\frac{A\sin^2\theta}{\Sigma}\left(d\phi-\Omega dt\right)^2+\frac{\Sigma}{\Delta}dr^2+\Sigma d\theta^2,
\end{equation}
where
\begin{equation}
\Omega=\frac{fHa}{A}.
\end{equation}
In this spacetime, the locations of horizons are determined by $\Delta(r_\Hor)=0$, and the angular velocity at the outer horizon $r_\Hor^+$ takes the form, 
\begin{equation}
\Omega_{\rm H}=\Omega\big|_{r=r_\Hor^+}=\frac{a}{fH}\bigg|_{r=r_\Hor^+}.\label{angvelathor}
\end{equation}
In the next  subsection, we shall proceed with the computation of the mass and angular momentum of rotating regular black holes.

\subsection{Komar's conserved quantity}
Here we use Komar's conserved quantity for the calculation of mass and angular momentum.
In the case of asymptotically flat spacetime, the Komar conserved quantity equals the ADM one, thus enabling us to obtain the ADM mass and angular momentum.
Komar's conserved quantity takes~\cite{Jeffrey:1984,Modak:2010fn} the form,
\begin{equation}
16\pi I=\int_{\partial V} {^*d\xi},
\end{equation}
where ${^*d\xi}$ denotes dual to a two-form $d\xi$, $\xi$ is Killing one-form, and the integration is extended over a spacelike surface $\partial V$ of the background spacetime depicted by Eq.~(\ref{rotating_metric}).
The Killing vector field related to mass is time-like, $\xi_{(t)}$, while that related to angular momentum is space-like, $\xi_{(\phi)}$.
We shall perform the calculations for mass and angular momentum.
We note that the metric contains coupling terms of $dt$ and $d\phi$, see Eq.~(\ref{rotating_metric}),  which makes it difficult to compute derivatives in the $(t,r,\theta,\phi)$ coordinates. Therefore, we introduce an orthonormal frame of one-forms for the derivations in the following subsections,
\begin{subequations}\label{orthoframe}
\begin{equation}
X_{(0)}=-\left(\frac{\Delta \Sigma}{A}\right)^{{1}/{2}}dt,
\end{equation}
\begin{equation}
X_{(1)}=\left(\frac{\Sigma}{\Delta}\right)^{{1}/{2}}dr,
\end{equation}
\begin{equation}
X_{(2)}=\Sigma^{{1}/{2}}d\theta,
\end{equation}
\begin{equation}
X_{(3)}=\left(\frac{A\sin^2\theta}{\Sigma}\right)^{{1}/{2}}\left(d\phi-\Omega dt\right).
\end{equation}
\end{subequations}

\subsubsection{Mass}
The mass can be precisely defined by the Komar conserved quantity $I_{(t)}$ of the time-like Killing vector field $\xi_{(t)}$. 
Specifically, it is determined as follows:
\begin{equation}\label{def:M}
M_{\rm K}=\lim_{r\rightarrow\infty}2I_{(t)}=\lim_{r\rightarrow\infty}\frac{1}{8\pi}\int_{\partial V}{}^*d\xi_{(t)},
\end{equation}
where the time-like Killing vector field is given with the help of the orthonormal frame Eq.~(\ref{orthoframe}), 
\begin{equation}
\xi_{(t)}=g_{\mu t}dx^\mu=\left(\frac{\Delta\Sigma}{A}\right)^{{1}/{2}}X_{(0)}-\left(\frac{A\sin^2\theta}{\Sigma}\right)^{{1}/{2}}\Omega X_{(3)}.
\end{equation}
Differentiating $\xi_{(t)}$ and taking its dual, we obtain
\begin{equation}
{^*d\xi_{(t)}}=\alpha X_{(2)}\wedge X_{(3)}-\beta X_{(1)}\wedge X_{(3)}+\gamma X_{(2)}\wedge X_{(0)}-\delta X_{(1)}\wedge X_{(0)},
\end{equation}
where
\begin{subequations}
\begin{equation}
\alpha=\frac{A^{{1}/{2}}}{\Sigma}\frac{\partial}{\partial r}\left(\frac{\Delta\Sigma}{A}\right)-\frac{A^{{3}/{2}}\Omega\sin^2\theta}{\Sigma^2}\frac{\partial}{\partial r}\Omega,
\end{equation}
\begin{equation}
\beta=\frac{1}{\Sigma}\left(\frac{A}{\Delta}\right)^{{1}/{2}}\frac{\partial}{\partial\theta}\left(\frac{\Delta\Sigma}{A}\right)-\frac{A\Omega\sin^2\theta}{\Sigma^2}\left(\frac{A}{\Delta}\right)^{{1}/{2}}\frac{\partial}{\partial\theta}\Omega,
\end{equation}
\begin{equation}
\gamma=-\left(\frac{\Delta\sin^2\theta}{A}\right)^{{1}/{2}}\frac{\partial}{\partial r}\left(\frac{A\Omega}{\Sigma}\right),
\end{equation}
\begin{equation}
\delta=-\left(\frac{1}{A\sin^2\theta}\right)^{{1}/{2}}\frac{\partial}{\partial\theta}\left(\frac{A\Omega\sin^2\theta}{\Sigma}\right).
\end{equation}
\end{subequations}

To perform integration in the original coordinates of $(t,r,\theta,\phi)$, we must convert ${^*d\xi_{(t)}}$ back to the formulation as follows:
\begin{equation}
{^*d\xi_{(t)}}=\tilde\alpha \dif r\wedge\dif t+\tilde\delta \dif\theta\wedge\dif\phi+\tilde\gamma\dif r\wedge\dif\phi+\tilde\beta\dif\theta\wedge\dif t,
\end{equation}
where
\begin{subequations}
\begin{equation}
\tilde\alpha=\beta\Omega\left(\frac{A\sin^2\theta}{\Delta}\right)^{{1}/{2}}+\delta\Sigma\left(\frac{1}{A}\right)^{{1}/{2}},
\end{equation}
\begin{equation}
\tilde\delta=\alpha\left(A\sin^2\theta\right)^{{1}/{2}},
\end{equation}
\begin{equation}
\tilde\gamma=-\beta\left(\frac{A\sin^2\theta}{\Delta}\right)^{{1}/{2}},
\end{equation}
\begin{equation}
\tilde\beta=-\gamma\left(\frac{\Delta}{A}\right)^{{1}/{2}}\Sigma-\alpha(A\sin^2\theta)^{{1}/{2}}\Omega.
\end{equation}
\end{subequations}
Here $t$ and $r$ are constants since we are calculating the mass in a two-dimensional sphere over simultaneous events.
Therefore, the Komar conserved quantity associated with  $\xi_{(t)}$ takes the form, 
\begin{equation}
I_{(t)}=\frac{1}{16\pi}\int\int\alpha\left(A\sin^2\theta\right)^{{1}/{2}}d\theta d\phi.
\end{equation}
After a tedious integral calculation, we obtain
\begin{equation}
I_{(t)}=\frac{1}{8} \left(H+a^2\right) \left[\frac{2 H F'+(F-1) H'}{a \sqrt{H}}\,{\tan^{-1}}\left(\frac{a}{\sqrt{H}}\right)-\frac{(F-1) H'}{H+a^2}\right],
\end{equation}
where a prime stands for derivative with respect to $r$.
According to the definition of mass, see Eq.~\eqref{def:M}, and the asymptotic behaviors, see Eqs.~\eqref{eq:series F} and \eqref{eq:series H}, we finally derive the mass,
\begin{equation}\label{eq:M}
M_{\rm K}=\lim_{r\to\infty}\frac{1}{4}(1-F)H'=-\lim_{r\to\infty}\frac{1}{4}\,\sum_{n=1,m=0}^\infty a_n b_m (2-m)r^{1-m-n}=-\frac{a_1}{2}.
\end{equation}
It is obvious that the Komar mass corresponds~\cite{Balart:2014cga,Ayon-Beato:1998hmi,Hayward:2005gi,Nicolini:2005vd,Brahma:2020eos} to the mass parameter $M$ selected in the metrics for the majority of rotating regular black holes. 
For convenience, we shall omit the subscript "K" and express it simply as $M$ in subsequent discussions.

\subsubsection{Angular momentum}
Similar to the definition of mass, the angular momentum is defined by
\begin{equation}\label{def:J}
    J_{\rm K}=-\lim_{r\rightarrow\infty}I_{(\phi)}=-\lim_{r\rightarrow\infty}\frac{1}{16\pi}\int_{\partial V}{}^*d\xi_{(\phi)},
\end{equation}
where the space-like Killing vector field $\xi_{(\phi)}$ takes the form in the orthonormal frame Eq.~(\ref{orthoframe}), 
\begin{equation}
\xi_{(\phi)}=g_{\mu \phi}dx^\mu=\left(\frac{A\sin^2\theta}{\Sigma}\right)^{{1}/{2}} X_{(3)}.
\end{equation}
We then write the dual of $\xi_{(\phi)}$,
\begin{equation}
{^*d\xi_{(\phi)}}=\hat\alpha X_{(2)}\wedge X_{(3)}-\hat\beta X_{(1)}\wedge X_{(3)}+\hat\gamma X_{(2)}\wedge X_{(0)}-\hat\delta X_{(1)}\wedge X_{(0)},
\end{equation}
where 
\begin{subequations}
\begin{equation}
\hat\alpha=\frac{A^{{3}/{2}}\sin^2\theta}{\Sigma^2}\frac{\partial}{\partial r}\Omega,
\end{equation}
\begin{equation}
\hat\beta=\frac{A\sin^2\theta}{\Sigma^2}\left(\frac{A}{\Delta}\right)^{{1}/{2}}\frac{\partial}{\partial\theta}\Omega,
\end{equation}
\begin{equation}
\hat\gamma=\left(\frac{\Delta}{A\sin^2\theta}\right)^{{1}/{2}}\frac{\partial}{\partial r}\left(\frac{A\sin^2\theta}{\Sigma}\right),
\end{equation}
\begin{equation}
\hat\delta=\left(\frac{1}{A\sin^2\theta}\right)^{{1}/{2}}\frac{\partial}{\partial\theta}\left(\frac{A\sin^2\theta}{\Sigma}\right).
\end{equation}
\end{subequations}
Here $t$ and $r$ are also constants because we are calculating the angular momentum in a two-dimensional sphere over simultaneous events.
Therefore, the Komar conserved quantity associated with $\xi_{(\phi)}$ is 
\begin{equation}
I_\phi=\frac{1}{16\pi}\int_{\partial V}{}^*d\xi_\phi=\frac{1}{16\pi}\int\int\hat\alpha\left(A\sin^2\theta\right)^{{1}/{2}}d\theta d\phi,
\end{equation}
and the integration gives
\begin{eqnarray}
\begin{split}
I_\phi 
&=\frac{1}{8 a^2 \sqrt{H(r)}}\Bigg\{-2F'(r) H(r) \left(a^2+H(r)\right)  \left[\left(a^2+H(r)\right)\tan ^{-1}\left(\frac{a}{\sqrt{H(r)}}\right)-a \sqrt{H(r)}\right] \\
&\;\;\;\;\; -[F(r)-1] H'(r) \left[\left(a^4 +2 a^2 H(r) +[H(r)]^2\right) \tan ^{-1}\left(\frac{a}{\sqrt{H(r)}}\right)-3 a^3 \sqrt{H(r)}-a [H(r)]^{3/2}\right]\Bigg\}.
\end{split}
\end{eqnarray}
According to the asymptotic behaviors, see Eqs.~\eqref{eq:series F} and \eqref{eq:series H}, we derive the angular momentum, 
\begin{equation}
\begin{split}
J=-\lim_{r\to\infty}I_\phi =-\lim_{r\to\infty}\frac{a}{8}\left[(F-1)H'-2F'(H+a^2)\right]=-\frac{aa_1}{2}=Ma,
\end{split}
\end{equation}
where the subscript "K" has been omitted as explained in the case of mass. 

We derive two physical quantities, mass and angular momentum, associated with a general rotating regular black hole. 
Further, we shall delve into the characteristic quantities involved in the process of a neutral scalar particle's incidence in the background spacetime of the rotating regular black holes described by Eq.~(\ref{rotating_metric}), and examine the particle's impact on the aforementioned physical quantities.

\section{Neutral massive scalar fields}\label{sec:scalar}
\subsection{Scalar field equation and flux}
The action of a complex scalar field $\Phi(t, r, \theta, \phi)$ in a general spacetime reads
\begin{equation}\label{eq: action}
S_\Phi=-\frac{1}{2}\int\dif^4x\sqrt{-g}\left[\partial_\nu\Phi\partial^\nu\Phi^*+(\mu^2+\Xi R)\Phi\Phi^*\right],
\end{equation}
where $\mu$ denotes the mass of scalar fields, $R$ the curvature, and $\Xi$ the non-minimal coupling constant, and $\sqrt{-g}$ takes the form,
\begin{equation}
\sqrt{-g}=\sqrt{-\rm{det}\,g_{\mu\nu}}=\Sigma\sin\theta,\label{gdeter}
\end{equation}
where Eq.~(\ref{rotating_metric}) has been used. By using the principle of least action, we derive the equation of motion,
\begin{equation}
\nabla_\nu\nabla^\nu\Phi=(\mu^2+\Xi R)\Phi.
\end{equation}

In order to separate variables in the spacetime of rotating regular black holes, we make the assumption,
\begin{equation}
\Phi(t, r, \theta, \phi)={\rm e}^{-i\omega t+im\phi}\mathscr{S}(\theta)\mathscr{R}(r),\label{sepvar}
\end{equation}
and then obtain the equations that govern $\mathscr{S}(\theta)$ and $\mathscr{R}(r)$, respectively,
\begin{equation}\label{S(theta)}
\frac{1}{\sin\theta}\frac{{\rm d}}{{\rm d}\theta}\left[\sin\theta\frac{{\rm d}}{{\rm d}\theta}\mathscr{S}(\theta)\right]+\left[a^2(\omega^2-\mu^2-\Xi R)\cos^2\theta-\frac{m^2}{\sin^2\theta}+\lambda\right]\mathscr{S}(\theta)=0,
\end{equation}
and
\begin{equation}\label{R(r)}
\Delta\frac{{\rm d}}{{\rm d}r}\left[\Delta \frac{{\rm d}}{{\rm d}r}\mathscr{R}(r)\right]+\left[\omega^2(H+a^2)^2-2afHm\omega+a^2m^2-\Delta(\mu^2H+\Xi RH+\lambda+a^2\omega^2)\right]\mathscr{R}(r)=0,
\end{equation}
where $\omega$ is the frequency of massive scalar fields, $m$ the azimuthal number with respect to the rotation axis, and $\lambda$ the separation parameter which can be fixed approximately as an eigenvalue of Eq.~(\ref{S(theta)}).

Introducing a tortoise coordinate in an outer horizon limit, 
\begin{equation}
\frac{\dif r_*}{\dif r}=\frac{H+a^2}{\Delta},
\end{equation}
we rewrite the radial equation as follows:
\begin{equation}
\frac{\dif^2}{\dif r_*^2}\mathscr{R}+(\omega-m\Omega_{\rm H})^2\mathscr{R}=0.
\end{equation}
Therefore, we obtain the solutions in the close proximity to outer horizons,
\begin{equation}
\mathscr{R}\sim {\rm e}^{\pm i(\omega-m\Omega_{\rm H})r_*},
\end{equation}
and take the negative sign as an ingoing wave.

The alterations in energy and angular momentum induced by complex scalar particles can be derived from the fluxes of energy and angular momentum related to the particles.
Once a complex scalar particle crosses an outer event horizon of a black hole, it becomes indistinguishable from the black hole. 
Hence, the alterations in mass and angular momentum related to black holes are closely linked to the scalar field fluxes at an outer horizon of black holes.
The fluxes can be derived from the energy-momentum tensor, which is obtained through the Lagrangian Eq.~\eqref{eq: action} of scalar fields.
Specifically, the energy-momentum tensor reads 
\begin{equation}
\begin{split}
T^\mu_\nu &=\sum_i\frac{\partial\mathscr{L}}{\partial(\partial_\mu\Phi^i)}\partial_\nu\Phi^i-\delta^\mu_\nu\mathscr{L}\\
&=\frac{1}{2}\partial^\mu\Phi\partial_\nu\Phi^*+\frac{1}{2}\partial^\mu\Phi^*\partial_\nu\Phi-\delta^\mu_\nu\left[\frac{1}{2}\partial_\mu\Phi\partial^\mu\Phi^*-\frac{1}{2}(\mu^2+\Xi R)\Phi\Phi^*\right],
\end{split}
\end{equation}
and the correlated energy flux of scalar fields at an outer horizon takes the form,
\begin{equation}\label{eq:flux_ene}
j_{t({\rm H})}=\frac{\dif E}{\dif t}=\lim_{r\to r^+_{\rm H}}\int{T^r_t\sqrt{-g}\dif\theta\dif\phi}=\omega(\omega-m\Omega_{\rm H})(H+a^2),
\end{equation}
where the normalization condition of $\mathscr{S}(\theta)$,
\begin{equation}
\int{\mathscr{S}(\theta)\mathscr{S}^*(\theta)\sin\theta\dif\theta\dif\phi}=1,
\end{equation}
has been used.
Similarly, the angular momentum flux of scalar fields at an outer horizon reads
\begin{equation}\label{eq:flux_ang}
j_{\phi({\rm H})}=\frac{\dif L}{\dif t}=-\lim_{r\to r^+_{\rm H}}\int{T^r_\phi\sqrt{-g}\dif\theta\dif\phi}=m(\omega-m\Omega_{\rm H})(H+a^2).
\end{equation}
The energy flux and angular momentum flux are crucial in verifying the second and third laws of black hole thermodynamics. The reason is that the fluxes represent how the energy and angular momentum flowing into black holes change with time. 
In addition, an important quantity closely related to black hole thermodynamics is the Hawking temperature.
\subsection{Hawking temperature}
We employ the methodology proposed in Ref.~\cite{Murata:2006pt} to compute the Hawking temperature of rotating regular black holes. 
The key step lies in reducing a 4-dimensional metric to a 2-dimensional one near an outer horizon, which is primarily accomplished by the reformulation of action Eq.~\eqref{eq: action}.
We rewrite the action by using Eq.~(\ref{gdeter}),
\begin{equation}
\begin{split}
S_\Phi 
&=-\frac{1}{2}\int{\dif^4x}\ \Sigma\sin\theta\Big[-\frac{(H+a^2)^2-\Delta a^2\sin^2\theta}{\Delta\Sigma} \partial_t\Phi\partial_t\Phi^*-\frac{a(H+a^2-\Delta)}{\Delta\Sigma}\left(\partial_t\Phi\partial_\phi\Phi^*+\partial_\phi\Phi\partial_t\Phi^*\right)\\
&\quad+\frac{\Delta-a^2\sin^2\theta}{\Delta\Sigma\sin^2\theta}\partial_\phi\Phi\partial_\phi\Phi^*+\frac{\Delta}{\Sigma}\partial_r\Phi\partial_r\Phi^*+\frac{1}{\Sigma}\partial_\theta\Phi\partial_\theta\Phi^*+(\mu^2+\Xi R)\Phi\Phi^*\Big],
\end{split}
\end{equation}
and then obtain its form in the vicinity of outer horizons by maintaining dominant terms, 
\begin{equation}
\begin{split}
S_\Phi\left[r\rightarrow r_{\rm H}^+\right]
&=\frac{1}{2}\int{\dif^4 x}\ \sin\theta\frac{a^2}{\Omega_{\rm H}^2\Delta}\Bigg[\partial_t\Phi\partial_t\Phi^*+\Omega_{\rm H}(\partial_t\Phi\partial_\phi\Phi^*+\partial_t\Phi^*\partial_\phi\Phi)\\
&\;\;\;\;\;+\Omega_{\rm H}^2\partial_\phi\Phi\partial_\phi\Phi^*-\frac{\Delta^2\Omega_{\rm H}^2}{a^2}\partial_r\Phi\partial_r\Phi^*\Bigg].\\
\end{split}
\end{equation}
Further, using the locally non-rotating coordinate, 
\begin{equation}
	\psi=\phi-\Omega_{\rm H}t,\qquad \tilde t=t,
\end{equation}
we transform the above form to be
\begin{equation}
S_\Phi\left[r\rightarrow r_{\rm H}^+\right]=-\frac{1}{2}\int{\dif^4 x}\sin\theta\frac{a}{\Omega_{\rm H}}\Big[-\frac{1}{f_e(r)}\partial_{\tilde t}\Phi\partial_{\tilde t}\Phi^*+f_e(r)\partial_r\Phi\partial_r\Phi^*\Big],
\end{equation}
where $f_e(r)$ is defined by
\begin{equation}
f_e(r)\equiv \frac{\Omega_{\rm H}\Delta}{a}.
\end{equation}
As a result, we reduce  the 4-dimensional action to a 2-dimensional one near an outer horizon, where the effective 2-dimensional metric is given by
\begin{equation}
\dif s^2=-f_e(r)\dif\tilde{t}^2+\frac{1}{f_e(r)}\dif r^2,
\end{equation}
and calculate the Hawking temperature according to Ref.~\cite{Murata:2006pt}, 
\begin{equation}\label{eq:Hawking_T}
T=\frac{1}{4\pi}\partial_r f_e\Big|_{r=r_{\rm H}^+}=\frac{\Omega_{\rm H}}{4\pi a}\partial_r\Delta\Big|_{r=r_{\rm H}^+}.
\end{equation}

We have provided all the necessary quantities except entropy for constructing the laws of thermodynamics. 
Next, we shall analyze the conditions required for the first, second and third laws of thermodynamics, respectively.

\section{Thermodynamics of rotating regular black holes under incidence of neutral scalar fields}\label{sec:thermo}
With a neutral scalar particle's incidence, the parameters of black holes undergo a transformation from $(M,J,r_{\rm H})$ to $(M+\dif M,J+\dif J,r_{\rm H}+\dif r_{\rm H})$.
In this process, we assume the perpetual existence of black hole event horizons, thereby ensuring that the condition $\Delta=0$ is always satisfied.
Specifically, the necessary conditions for the existence of event horizons is
\begin{equation}\label{eq:horizon}
\Delta(M,J,r_{\rm H})=F(M,r_{\rm H})H(M,r_{\rm H})+\frac{J^2}{M^2}=0,
\end{equation}
where we have rewritten $F(r)$ and $H(r)$ as $F(M,r)$ and $H(M,r)$, respectively, since $M$ is parameter of these functions as shown in Eq.~\eqref{eq:M}. 
However, the angular momentum $J$ is not a parameter of $F$ and $H$ because they are defined in a static and spherically symmetric spacetime and independent of any rotation introduced later.
After a neutral scalar particle's incidence, the condition Eq.~(\ref{eq:horizon}) changes to
\begin{equation}
\Delta(M+\dif M,J+\dif J,r_{\rm H}+\dif r_{\rm H})=\frac{\partial\Delta}{\partial M}\Big|_{r=r_{\rm H}}\dif M+\frac{\partial\Delta}{\partial J}\Big|_{r=r_{\rm H}}\dif J+\frac{\partial\Delta}{\partial r}\Big|_{r=r_{\rm H}}\dif r_{\rm H}=0,
\end{equation}
where
\begin{eqnarray}
\frac{\partial\Delta}{\partial M}\Big|_{r=r_{\rm H}}&=&-\frac{1}{H}\left(a^2\frac{\partial H}{\partial M}-H^2\frac{\partial F}{\partial M}+2H\frac{J^2}{M^3}\right)\Big|_{r=r_{\rm H}},\\
\frac{\partial\Delta}{\partial J}\Big|_{r=r_{\rm H}}&=&\frac{2J}{M^2},\\
\frac{\partial\Delta}{\partial r}\Big|_{r=r_{\rm H}}&=&-\frac{1}{H}\left(a^2\frac{\partial H}{\partial r}-H^2\frac{\partial F}{\partial r}\right)\Big|_{r=r_{\rm H}}.
\end{eqnarray}
Thus, we establish the correlation among $\dif r_{\rm H}$, $\dif M$, and $\dif J$ from the above four equations,
\begin{equation}\label{eq:r_M_J}
\dif r_\Hor=-\frac{\partial\Delta}{\partial M}\left(\frac{\partial\Delta}{\partial r}\right)^{-1}\Big|_{r=r_{\rm H}}\dif M-\frac{\partial\Delta}{\partial J}\left(\frac{\partial\Delta}{\partial r}\right)^{-1}\Big|_{r=r_{\rm H}}\dif J,
\end{equation}
with which we further construct the laws of thermodynamics.

\subsection{First law of thermodynamics}\label{subsec:the first law}
The differential form of the first law of thermodynamics for black holes establishes a relationship among the first-order differentials of physical quantities. 
As neutral scalar particles induce changes in mass and angular momentum, the first law of thermodynamics should describe the correlation among first-order differentials of entropy, mass, and angular momentum.
A significant thermodynamic quantity is the entropy of a black hole. 
As adopted in the thermodynamics of singular black holes, we define the entropy of a regular black hole by using the  Bekenstein-Hawking entropy, 
\begin{equation}
S_{\rm B\Hor}=\frac{1}{4}\mathscr{A}_\Hor,\label{BHAT}
\end{equation}
and compute the area of out horizons by using Eq.~(\ref{rotating_metric}),
\begin{equation}
\mathscr{A}_\Hor=\int{\sqrt{g_{\theta\theta}g_{\phi\phi}}\dif\theta\dif\phi}\Big|_{r=r_\Hor}=4\pi(H+a^2)\Big|_{r=r_\Hor}.
\end{equation}
When the black hole parameters change, the associated entropy undergoes a corresponding change,
\begin{equation}
\dif S_{\rm B\Hor}=\frac{\partial S_{\rm B\Hor}}{\partial M}\dif M+\frac{\partial S_{\rm B\Hor}}{\partial J}\dif J+\frac{\partial S_{\rm B\Hor}}{\partial r_\Hor}\dif r_\Hor.
\end{equation}
After utilizing the relationship Eq.~\eqref{eq:r_M_J}, we derive
\begin{equation}\label{TFirst}
    \dif S_{\rm B\Hor}=\hat{A}\dif M+\hat{B}\dif J,
\end{equation}
where
\begin{subequations}
    \begin{equation}
        \hat{A}=-\frac{1}{4TH(H+a^2)}\left\{H^2\left(\frac{\partial H}{\partial r_\Hor}\frac{\partial F}{\partial M}-\frac{\partial H}{\partial M}\frac{\partial F}{\partial r_\Hor}\right)+\frac{2a^2}{M}\left[H^2\frac{\partial F}{\partial r_\Hor}-(H+a^2)\frac{\partial H}{\partial r_\Hor}\right]\right\},\label{ahat}
    \end{equation}
    \begin{equation}
        \hat{B}=\frac{a}{2TMH(H+a^2)}\left[H^2\frac{\partial F}{\partial r_\Hor}-(H+a^2)\frac{\partial H}{\partial r_\Hor}\right],\label{bhat}
    \end{equation}
\end{subequations}
and the following relationship has been used,
\begin{equation}
    4\pi TS_{\rm B\Hor}=\frac{\partial \Delta}{\partial r_\Hor}=-\frac{1}{H}\left(a^2\frac{\partial H}{\partial r_\Hor}-H^2\frac{\partial F}{\partial r_\Hor}\right).
\end{equation}

In the thermodynamics of singular black holes, the first law takes the form,
\begin{equation}
    \dif M=T\dif S_{\rm B\Hor}+\Omega_\Hor\dif J.\label{fltnrbh}
\end{equation}
When comparing Eq.~(\ref{fltnrbh}) with Eq.~(\ref{TFirst}),  we give $\hat{A}$ and $\hat{B}$ if the first law maintains unchanged for regular black holes,
\begin{subequations}
    \begin{equation}\label{eq:1_A}
    \frac{1}{\hat{A}}=T,
    \end{equation}
    \begin{equation}\label{eq:B_A}
    \quad -\frac{\hat{B}}{\hat{A}}=\Omega_\Hor.
    \end{equation}
\end{subequations}
Therefore, we determine the necessary conditions by using Eqs.~(\ref{ahat}) and (\ref{bhat}) for a rotating regular black hole to satisfy the first law of thermodynamics,
\begin{subequations}
    \begin{equation}\label{Fcon_1}
        \frac{\partial H}{\partial r_\Hor}\frac{\partial F}{\partial M}-\frac{\partial H}{\partial M}\frac{\partial F}{\partial r_\Hor}=-4,
    \end{equation}
    \begin{equation}\label{Fcon_2}
        H\frac{\partial F}{\partial r_\Hor}-(1-F)\frac{\partial H}{\partial r_\Hor}=-2M.
    \end{equation}
\end{subequations}
\subsection{Second law of thermodynamics}\label{subsec:the second law}
The second law of black hole thermodynamics dictates that the entropy of a system can only increase or remain constant during the system's evolution without external interaction.
This evolution naturally includes a scalar particle's incidence into a black hole. 
Therefore, by utilizing Eqs.~\eqref{eq:flux_ene} and ~\eqref{eq:flux_ang}, we derive the temporal variations of both energy and angular momentum of black holes,
\begin{subequations}
    \begin{equation}\label{eq:dM_dt}
        \dif M= \frac{\dif E}{\dif t}\Big|_{r=r_{\Hor}}\dif t=\omega(\omega-m\Omega_\Hor)(H+a^2)\dif t,
    \end{equation}
    \begin{equation}\label{eq:dJ_dt}
        \dif J= \frac{\dif L}{\dif t}\Big|_{r=r_{\Hor}}\dif t=m(\omega-m\Omega_\Hor)(H+a^2) \dif t.
    \end{equation}
\end{subequations}
Thus, we obtain the relationship between $ \dif  S_{\rm B\Hor}$ and $\dif t$,
\begin{equation}
\label{eq:S_t}
    \dif  S_{\rm B\Hor}=\hat{A}\dif M+\hat{B}\dif J=(\hat{A}\omega+\hat{B}m)(\omega-m\Omega_\Hor)(H+a^2)\dif t.
\end{equation}
Moreover, considering that frequency $\omega$ and integer $m$ are arbitrary, we deduce the necessary and sufficient conditions for $\dif  S_{\rm B\Hor}\geq 0$,
\begin{equation}\label{eq:second_citi}
    \hat{A}>0 \qquad{\rm{and}} \qquad \frac{\hat{B}}{\hat{A}}=-\Omega_\Hor.
\end{equation}
Again using Eqs.~(\ref{ahat}) and (\ref{bhat}), we reformulate the necessary and sufficient conditions for a rotating regular black hole to satisfy the second law of thermodynamics,
\begin{subequations}
\begin{equation}\label{Scon_2}
\frac{\partial H}{\partial r_\Hor}\frac{\partial F}{\partial M}-\frac{\partial H}{\partial M}\frac{\partial F}{\partial r_\Hor}<0,
\end{equation}
\begin{equation}\label{Scon_1}
\left(H+\frac{M}{2}\frac{\partial H}{\partial M}\right)\frac{\partial F}{\partial r_\Hor}-\left(1-F+\frac{M}{2}\frac{\partial F}{\partial M}\right)\frac{\partial H}{\partial r_\Hor}=0.
\end{equation}
\end{subequations}

By comparing Eqs.~(\ref{Fcon_1}) and (\ref{Fcon_2}) with Eqs.~(\ref{Scon_2}) and (\ref{Scon_1}), we see that the validity of the first law indicates the validity of the second law. Or, according to Eqs.~\eqref{eq:1_A} and ~\eqref{eq:B_A}, we obtain that the first law gives rise to
\begin{equation}
	\dif S_{\rm B\Hor}=\frac{1}{T}(\omega-m\Omega_\Hor)^2(H+a^2)\dif t,
\end{equation}
which implies that the second law is naturally satisfied.
This result seems to be plausible but actually incorrect. The reason has two folds, one is that we have to compute the entropy of black holes and the entropy of scalar particles but not the entropy only for black holes, and the other fold is that a (rotating) regular black hole  does not satisfy the first law of a (rotating) singular black hole. 
In fact, the first law of mechanics for static and spherically symmetric regular black holes usually contains~\cite{Zhang:2016ilt} additional terms related to regularized parameters, which plays~\cite{Fan:2016hvf} a non-negligible role in constructing the first law of thermodynamics.
Further investigation is required to establish the first and second laws for (rotating) regular black holes.

In accordance with the above clarifications in this section, we note that both the first and second laws of thermodynamics are established on Bekenstein-Hawking's area theorem Eq.~(\ref{BHAT}). 
Therefore, if Bekenstein-Hawking's area theorem is violated, the two laws are also violated. Conversely, if the second law is violated, the entropy will decrease. This implies that the basis of Bekenstein-Hawking's area theorem, i.e., the entropy increases during evolution, is lost, meaning the nonexistence of the theorem.

\subsection{Third law of thermodynamics and weak cosmic censorship conjecture}\label{subsec:the third law}
The above discussions of thermodynamics are based on the general premise that an incident scalar particle does not destroy an event horizon of black holes.
In spacetime with singularities, this premise is also required by the weak cosmic censorship conjecture, where no naked singularities are allowed. 
However, the weak cosmic censorship conjecture loses its necessity in regular black holes, where no singularities exist. 
Therefore, we can comprehend the weak cosmic censorship conjecture as a mechanism that safeguards the thermodynamic stability of black holes because thermodynamic quantities are only well-defined on event horizons.
Next we are going to reexamine whether an incident scalar particle breaks an event horizon, i.e., to verify whether the weak cosmic censorship conjecture holds for a rotating regular black hole.
Since the energy and angular momentum of scalar particles are very small compared to those of black holes, incident scalar particles can only destroy an event horizon when a black hole is near its ultimate state.

In general, there exist two primary categories of ultimate states for a rotating regular black hole. 
The first scenario entails that a black hole initially possesses multiple horizons and these horizons subsequently coalesce into a solitary one with a non-zero radius in the final stage of black holes - this is referred to as an extreme black hole case.
The alternative scenario is that a black hole ultimately possesses one event horizon with a vanishing radius, which can be conceived of as a one-way wormhole featuring a null throat~\cite{Simpson:2018tsi,Brahma:2020eos}.

\subsubsection{In the first scenario}\label{subsubsec:first scenario}
Let us commence by scrutinizing the first scenario.
According to the existence condition Eq.~\eqref{eq:horizon} of horizons, we rewrite $r_\Hor$ as a function of $M$ and $J$,
\begin{equation}\label{eq:rh}
    r_\Hor=r_\Hor(M,J).
\end{equation}
In terms of $\lim_{r\to\infty}\Delta=r^2$, we know that the value of $\Delta$ is negative between an inner horizon and an outer one and that a minimum value of $\Delta$, denoted by $\Delta_\Min$, satisfies the condition in terms of Eq.~(\ref{defDelta}),
\begin{equation}\label{eq:dif_DeltaM}
        \frac{\partial\Delta(M,J,r_{\Min})}{\partial r_{\Min}}=F(M,r_{\Min})\frac{\partial H(M,r_{\Min})}{\partial r_{\Min}}+H(M,r_{\Min})\frac{\partial F(M,r_{\Min})}{\partial r_{\Min}}=0,
\end{equation}
from which we can deduce that $r_\Min$ only depends on $M$,
\begin{equation}\label{eq:rmin}
    r_\Min=r_\Min(M).
\end{equation}
If there are multiple minimum values, the largest one among them will be selected.
We note that the extreme horizon radius $r_e$ satisfies both Eq.~\eqref{eq:horizon} and  Eq.~\eqref{eq:dif_DeltaM}. 
As a result, we are able to express the angular momentum $J_e$ of an extreme black hole as a function of $M$,
\begin{equation}
  J_e=\sqrt{-M^2F(M,r_e(M))H(M,r_e(M))}\equiv j_e(M),\label{angmomextbh}
\end{equation}
where
\begin{equation}\label{eq:re}
    r_e(M)=r_\Min(M).
\end{equation}
Eq.~(\ref{angmomextbh}) shows that the upper bound on the angular momentum is determined only by mass, thus it is named as ``extreme function".
According to Eq.~\eqref{eq:Hawking_T}, the Hawking temperature of an extreme black hole is zero.
The third law of thermodynamics stipulates that it requires an infinite amount of time, or equivalently, an infinite number of steps to reach the absolute zero temperature.
This implies that the evolution of a black hole into its extreme configuration cannot be achieved through any finite number of steps.
Subsequently, we shall demonstrate that the weak cosmic censorship conjecture is compatible with the third law of thermodynamics in the first scenario.

From the above analyses the condition for existence of event horizons reads 
\begin{equation}
    \Delta_\Min\leq 0.
\end{equation}
By using Eqs.~(\ref{defDelta}) and (\ref{angmomextbh}), we rewrite it as follows:
\begin{equation}\label{eq:horizon_exists}
    J^2\leq -M^2F(M,r_\Min(M))H(M,r_\Min(M))\equiv j_e^2(M),
\end{equation}
where the equality is valid only for extreme black holes.
Therefore, for a near-extreme black hole we assume the relationship between angular momentum $J_{ne}$ and mass $M$,
\begin{equation}
    J_{ne}=(1-\epsilon)j_e(M),\label{amne}
\end{equation}
where $\epsilon$ is an arbitrary positive infinitesimal parameter, $0<\epsilon\ll 1$.
After a neutral scalar particle's incidence into a near-extreme black hole, when we consider Eqs.~(\ref{eq:dJ_dt}) and (\ref{amne}), the angular momentum becomes
\begin{equation}
    J_{ne}+dJ=(1-\epsilon)j_e(M)+m(\omega-m\Omega_{ne})(H+a^2)\dif t,\label{angmominc}
\end{equation}
and when we consider Eq.~(\ref{eq:dM_dt}), the extreme function then reads
\begin{equation}
    j_e(M+\dif M)=j_e(M)+\frac{\dif j_e(M)}{\dif M}\omega(\omega-m\Omega_{ne})(H+a^2)\dif t.\label{extfuninc}
\end{equation}

Next we shall prove that the weak cosmic censorship conjecture is consist with the third law of thermodynamics in the first scenario, that is, both are valid, or both are broken. According to Eq.~\eqref{eq:horizon_exists}, the condition for absence of horizons after a neutral scalar particle's incidence takes the form,
\begin{equation}
    J_{ne}+\dif J>j_e(M+\dif M),\label{angmomdif}
\end{equation}
which implies  invalidity of the weak cosmic censorship conjecture, and the condition for a black hole to reach its extreme configuration is
\begin{equation}
    J_{ne}+\dif J=j_e(M+\dif M),\label{angmomdifext}
\end{equation}
which implies invalidity of the third law of thermodynamics.
More specifically, considering Eqs.~(\ref{angmominc}) and (\ref{extfuninc}) we combine Eqs.~(\ref{angmomdif}) and (\ref{angmomdifext}) as
\begin{equation}\label{ineq:des_con}
    -\epsilon j_e(M)\geq\left(\frac{\dif j_e(M)}{\dif M}\omega-m\right)(\omega-m\Omega_{ne})(H+a^2)\dif t,
\end{equation}
where the equality holds for an extreme configuration. From Eqs.~(\ref{eq:dM_dt}) and (\ref{eq:dJ_dt}) we deduce $ \frac{\dif j_e(M)}{\dif M}>0$.
Therefore, we determine that the right-hand side of Eq.~\eqref{ineq:des_con} satisfies\footnote{For a function, $f(x)=(x-a)(x-b)$, it takes its minimum, $-(a-b)^2/4$, when $x=(a+b)/2$.} the following inequality,
\begin{eqnarray}\label{ineq:des_con_2}
\left(\frac{\dif j_e(M)}{\dif M}\omega-m\right)(\omega-m\Omega_{ne})(H+a^2)\dif t&=&m^2\frac{\dif j_e(M)}{\dif M}\left[\frac{\omega}{m}-\left(\frac{\dif j_e(M)}{\dif M}\right)^{-1}\right]\left(\frac{\omega}{m}-\Omega_{ne}\right)(H+a^2)\dif t\nonumber \\
&\geq& -\frac{m^2}{4}\frac{\dif j_e(M)}{\dif M}\left[\left(\frac{\dif j_e(M)}{\dif M}\right)^{-1}-\Omega_{ne}\right]^2(H+a^2)\dif t,
\end{eqnarray}
where the equality is valid for the condition: $\frac{\omega}{m}=\frac{1}{2}\left[\left(\frac{\dif j_e(M)}{\dif M}\right)^{-1}+\Omega_{ne}\right]$.
In order to simplify Eq.~(\ref{ineq:des_con_2}), we expand $\Omega_{ne}(M,J_{ne},r_{ne})$ near the extreme configuration $\epsilon=0$ as
\begin{equation}
\Omega_{ne}(M,J_{ne},r_{ne})=\Omega_e-j_e(M)\kappa\epsilon+O(\epsilon^2),\label{omeganeexp}
\end{equation}
where $\kappa\equiv \frac{\dif \Omega_\Hor(M,J,r_\Hor)}{\dif J}\Big|_{J=j_e(M)}$.
Considering Eqs.~\eqref{ineq:des_con_2} and (\ref{omeganeexp}), we refine Eq.~\eqref{ineq:des_con} to be
\begin{equation}\label{ineq:des_con_3}
    \epsilon \leq \frac{m^2}{4j_e(M)}\frac{\dif j_e(M)}{\dif M}\left[\left(\frac{\dif j_e(M)}{\dif M}\right)^{-1}-\Omega_{e}+j_e(M)\kappa\epsilon\right]^2(H+a^2)\dif t.
\end{equation}
\begin{itemize}
\item 
If we have the condition, 
\begin{equation}
\left(\frac{\dif j_e(M)}{\dif M}\right)^{-1}-\Omega_{e}=0,\label{sufnecconfwt}
\end{equation} 
the above inequality becomes
\begin{equation}
     \epsilon \leq \frac{m^2}{4j_e(M)}\frac{\dif j_e(M)}{\dif M}\left[j_e(M)\right]^2\kappa^2(H+a^2)\epsilon^2\dif t.
\end{equation}
After comparing the orders of infinitesimals on both sides, we conclude that this inequality never holds, indicating that both the third law of thermodynamics and weak cosmic censorship conjecture are valid
in this scenario.
\item If we have the condition,  
\begin{equation}
\left(\frac{\dif j_e(M)}{\dif M}\right)^{-1}-\Omega_{e}\neq0, 
\end{equation}
we keep the finite term and ignore infinitesimals in Eq.~\eqref{ineq:des_con_3}, 
\begin{equation}
    \epsilon \leq \frac{m^2}{4j_e(M)}\frac{\dif j_e(M)}{\dif M}\left[\left(\frac{\dif j_e(M)}{\dif M}\right)^{-1}-\Omega_{e}\right]^2(H+a^2)\dif t.
\end{equation}
Because both $\epsilon$ and $\dif t$ are same order of infinitesimals, we can always find an appropriate time interval $\dif t$ such that this inequality holds, which means that both the third law of thermodynamics and weak cosmic censorship conjecture are broken.
\end{itemize}

\subsubsection{In the second scenario}\label{subsubsec:second scenario}
Now we examine the second scenario.
Considering Eq.~(\ref{angmomextbh}), we redefine $j_e(M)$ as follows:
\begin{equation}
     j_e(M)\equiv\sqrt{-M^2F(M,r)H(M,r)}\bigg|_{r=0}.
\end{equation}
According to Eq.~\eqref{eq:Hawking_T}, the Hawking temperature of this ultimate state is no longer zero.
Therefore, following the analyses made for the first scenario, we conclude that the third law holds, but the weak cosmic censorship conjecture may not hold in the second scenario because event horizons may disappear or black holes may transform into a wormhole.

\subsubsection{Results in the two scenarios}\label{subsubsec:two scenarios}
 
In the two scenarios Eq.~(\ref{sufnecconfwt}) is the sufficient and necessary condition for the weak cosmic censorship conjecture to be valid,
and it is also the necessary condition for the third law of thermodynamics to be valid in an extreme black hole. 
However, if the ultimate state of a black hole is a one-way wormhole in the second scenario, the weak cosmic censorship conjecture may not hold but the third law of thermodynamics still works. This gives rise to an interesting issue to establish the so-called wormhole thermodynamics to describe the evolutionary process from a black hole to a wormhole.

Moreover, we express Eq.~(\ref{sufnecconfwt}) in a specific form, i.e., in terms of $F(M, r)$ and $H(M,r)$. 
 By using Eqs.~(\ref{angmomextbh}) and (\ref{eq:dif_DeltaM}), we derive
\begin{equation}
	\frac{\dif j_e(M)}{\dif M}=\frac{j_e(M)}{M}-\frac{M^2}{2j_e(M)}\left(H(M,r_e)\frac{\partial F(M,r_e)}{\partial M}+F(M,r_e)\frac{\partial H(M,r_e)}{\partial M}\right),
\end{equation}
and then substituting the above equation into Eq.~(\ref{sufnecconfwt}) and considering Eqs.~(\ref{angvelathor}) and (\ref{eq:re}), we finally rewrite Eq.~(\ref{sufnecconfwt}) to be
\begin{equation}\label{Tcon_2}
H(M,r_e)\frac{\partial F(M,r_e)}{\partial M}+F(M,r_e)\frac{\partial H(M,r_e)}{\partial M}=-\frac{2H(M,r_e)}{M},
\end{equation}
where the definition of $r_e$ is given by Eqs.~\eqref{eq:dif_DeltaM}, \eqref{eq:rmin}, and ~\eqref{eq:re} in the first scenario, and  $r_e=0$ in the second scenario.
\subsection{Interpretation}
In the above three subsections, we have examined the conditions for the first, second, and third laws of thermodynamics to be valid, as well as for the weak cosmic censorship conjecture, see Eqs.~(\ref{Fcon_1}), (\ref{Fcon_2}), (\ref{Scon_2}),  (\ref{Scon_1}), and (\ref{Tcon_2}).
We note that the definition of $r_e$ given by Eqs.~\eqref{eq:dif_DeltaM}, ~\eqref{eq:rmin}, and ~\eqref{eq:re} is solely determined by metric functions of a static spacetime, thus rendering the validity condition Eq.~(\ref{Tcon_2}) for both the third law and weak cosmic censorship conjecture completely dependent on metric functions of a static spacetime.
However, in the validity conditions for the first and second laws, see Eqs.~(\ref{Fcon_1}), (\ref{Fcon_2}), (\ref{Scon_2}), and  (\ref{Scon_1}), we note that there exists a variable $r_\Hor$ that varies with the rotation parameter $a$ in addition to metric functions of a static spacetime.

When a static metric is given, the minimum horizon of a rotating black hole corresponds to $r_e$ of an extreme  configuration, where $r_e$ is independent of the rotation parameter $a$. Moreover, the maximum horizon $r_{max}$ also depends only on a static metric. 
Therefore, the range of horizons for a rotating black hole is $r\in[r_e,r_{\Hor(s)}]$, where $``(s)"$ denotes a static black hole regarded as a seed.
We now rewrite the validity conditions for the first law as
\begin{subequations}
    \begin{equation}\label{Fcon_1_4_4}
        \frac{\partial H}{\partial r}\frac{\partial F}{\partial M}-\frac{\partial H}{\partial M}\frac{\partial F}{\partial r}=-4,
    \end{equation}
    \begin{equation}\label{Fcon_2_4_4}
        H\frac{\partial F}{\partial r}-(1-F)\frac{\partial H}{\partial r}=-2M.
    \end{equation}
\end{subequations}
and for the second law as
\begin{subequations}
\begin{equation}\label{Scon_2_4_4}
     \frac{\partial H}{\partial r}\frac{\partial F}{\partial M}-\frac{\partial H}{\partial M}\frac{\partial F}{\partial r}<0,
\end{equation}
\begin{equation}\label{Scon_1_4_4}
      \left(H+\frac{M}{2}\frac{\partial H}{\partial M}\right)\frac{\partial F}{\partial r}-\left(1-F+\frac{M}{2}\frac{\partial F}{\partial M}\right)\frac{\partial H}{\partial r}=0,
\end{equation}
\end{subequations}
where $r\in[r_e,r_{\Hor(s)}]$.
In this way, we express the validity conditions only by metric functions $F$ and $H$ of a static seed spacetime. In other words, we determine the conditions under which a rotating black hole satisfies the laws of thermodynamics just by means of the metric functions of its static seed black hole.
We note that such conditions are valid for any rotating black holes constructed by the revised NJA since our calculations are made in the spacetime beyond event horizons.

\section{Application to three models}\label{sec:sp_model}
According to the number of shape functions that appear in a metric, we classify~\cite{Lan:2023cvz} regular black holes into two types, where these regular black holes are static and spherically symmetric and regarded as a seed of rotating regular black holes. 
In the first type,  $H(r)=r^2$ and only $F(r)$ is to be determined, see Eq.~(\ref{lineelemstatic}), which is called ``single function case", such as Bardeen black holes~\cite{Bardeen:1968nsg}, Hayward black holes~\cite{Hayward:2005gi}, noncommutative black holes~\cite{Nicolini:2005vd}, and the other widely studied black holes~\cite{Ayon-Beato:1998hmi,Balart:2014cga,Abdujabbarov:2016hnw}. 
In the second type, both $F(r)$ and $H(r)$ are to be determined, which is called ``double function case", such as black-bounce solutions~\cite{Simpson:2018tsi}, loop quantum gravity black holes~\cite{Brahma:2020eos}, and the other black holes that have recently attracted~\cite{Lobo:2020ffi} attentions and discussions.
Next we shall discuss the thermodynamics of these two types of black holes, respectively, 
where Hayward black holes, black-bounce black holes, and loop quantum gravity black holes are chosen as specific examples.
There are two reasons for our choices: One reason is that these three black holes correspond to the three categories of construction for regular black holes, which has been mentioned in Sec.~\ref{sec:intr};  the other is that these three black holes correspond to different  types of shape functions, from which we are able to gain a more comprehensive understanding of the thermodynamic laws for regular black holes.

\subsection{Single function case}
\subsubsection{Conditions}
In Eq.~(\ref{lineelemstatic}) $H(r)$ is fixed, $H(r)=r^2$, and thus the necessary conditions, Eqs.~\eqref{Fcon_1_4_4} and \eqref{Fcon_2_4_4}, for a rotating regular black hole to satisfy the first law of thermodynamics become
\begin{subequations}
    \begin{equation}\label{Fcon_1_5_1}
        \frac{\partial F(M,r)}{\partial M}=-\frac{2}{r},
    \end{equation}
    \begin{equation}\label{Fcon_2_5_1}
        r^2\frac{\partial F(M,r)}{\partial r}-2[1-F(M,r)]r+2M=0,
    \end{equation}
\end{subequations}
where $r\in[r_e,r_{\Hor(s)}]$.
From Eq.~\eqref{Fcon_1_5_1}, we solve $F(M,r)$ as follows:
\begin{equation}
    F(M,r)=1-\frac{2M}{r}+\sigma(r),\label{ofcF}
\end{equation}
where $\sigma(r)$ is a function independent of $M$.
Then substituting Eq.~\eqref{ofcF} into Eq.~\eqref{Fcon_2_5_1}, we obtain
\begin{equation}
    \sigma(r)=\frac{C_1}{r^2},
\end{equation}
where $C_1$ is integration constant.
As a result, we fix $F(M,r)$ in the single function case,
\begin{equation}\label{Fcon_3_1}
	F(M,r)=1-\frac{2M}{r}+\frac{C_1}{r^2},
\end{equation}
with which the rotating regular black hole of the single function case satisfies the same first law of thermodynamics as its seed black hole does.

From Eq.~(\ref{Fcon_3_1}) together with Eq.~(\ref{defDelta}) we give the horizon of rotating regular black holes, $r_\Hor=M+\sqrt{M^2-a^2-C_1}$, whose lower limit is $M$, corresponding to $a=\sqrt{M^2-C_1}$, and whose upper limit is $M+\sqrt{M^2-C_1}$, corresponding to $a=0$.
In order to ensure the existence of horizons and a non-vanishing rotation parameter, $C_1$ should satisfy $C_1<M^2$.
Therefore, it is worth noting that the range of $r$ is no longer from zero to infinity, but coincides with that of event horizons of rotating regular black holes, $[M, M+\sqrt{M^2-C_1})$.
Within this range, the rotating black hole is reduced to a Kerr black hole when $C_1=0$, and to a Kerr-Newman black hole when $C_1=Q^2$.
We have seen that the first law of thermodynamics imposes a highly stringent constraint on metrics, which indicates that a static seed black hole must have a metric with such an $F(M, r)$ as Eq.~\eqref{Fcon_3_1}. 
However, it is quite challenging to construct a continuous metric function $F(M,r)$ in the range of $r\in[0, \infty)$, where this function returns to Eq.~(\ref{Fcon_3_1}) when $r$ is fixed to the range of $[M, M+\sqrt{M^2-C_1})$.

Next, we examine the second law of thermodynamics.
The necessary conditions, Eqs.~\eqref{Scon_2_4_4} and \eqref{Scon_1_4_4}, become
\begin{subequations} 
    \begin{equation}\label{Scon_2_1}
        \frac{\partial F(M,r)}{\partial M}<0,
    \end{equation}
    \begin{equation}\label{Scon_2_2}
        r\frac{\partial F(M,r)}{\partial r}-M\frac{\partial F(M,r)}{\partial M}=2(1-F(M,r)),
    \end{equation}
\end{subequations}
where $r\in[r_e,r_{\Hor(s)}]$.
To verify the validity of the aforementioned conditions, we must rely on a specific metric function $F(M,r)$, which will be discussed in detail in the following models.

Finally, owing to $\Delta(M,r)\big|_{r=0}=a^2>0$, the ultimate state of black holes  is an extreme configuration in the single function case.
Therefore, for the third law of thermodynamics and weak cosmic censorship conjecture, the validity condition Eq.~(\ref{Tcon_2}) is reduced to be
\begin{equation}\label{Tcon_5_1}
    \frac{\partial F(M,r_e)}{\partial M}=-\frac{2}{M}.
\end{equation}
Since $r_e$ is a function of $M$, we cannot give $F(M, r)$ by a direct integration from the above equation, but leave it to specific models.

\subsubsection{Hayward model}
The specific model we choose here is the static and spherically symmetric Hayward black hole~\cite{Hayward:2005gi,Bambi:2013ufa}, whose shape function reads
\begin{equation}
    F_{\rm H}(r)=1-\frac{2Mr^2}{r^3+2L^2M},\label{samphaybh}
\end{equation}
where the regularization parameter $L$ is a convenient encoding of the central energy density $3/(8\pi L^2)$.
The field source of this black hole solution corresponds to the vacuum energy density distribution,
\begin{equation}
    \rho(r)=\frac{3L^2M^2}{2\pi(r^3+2L^2M)^2},
\end{equation}
and its radial pressure, $p_r$, and transverse pressures, $p_\theta$ and $p_\phi$, read
\begin{equation}
    p_r=-\frac{3L^2M^2}{2\pi(r^3+2L^2M)^2},
\end{equation}
\begin{equation}
    p_\theta=p_\phi=\frac{3(r^3-L^2M)L^2M^2}{\pi(r^3+2L^2M)^3}.
\end{equation}
If the first law of thermodynamics is required, when comparing Eq.~(\ref{samphaybh}) with Eq.~\eqref{Fcon_3_1}, we obtain the integration constant,
\begin{equation}\label{eq:Hayward_C1}
    C_1=\frac{4M^2L^2r}{r^3+2L^2M},
\end{equation}
where $r\in[M, M+\sqrt{M^2-C_1})$ and $C_1<M^2$.
It is obvious that the right-hand side of Eq.~\eqref{eq:Hayward_C1} is not a constant, so the first law of thermodynamics does not hold for rotating Hayward black holes.

For the second law of thermodynamics, Eq.~\eqref{Scon_2_1} is valid due to $r>0$,
\begin{equation}
    \frac{\partial F(M,r)}{\partial M}=-\frac{2r^5}{(r^3+2L^2M)^2}<0.
\end{equation}
However,  Eq.~\eqref{Scon_2_2} takes the form, 
\begin{equation}
    -\frac{16 L^2 M^2 r^2}{\left(2 L^2 M+r^3\right)^2}=0,
\end{equation}
which does not hold for $L\neq 0$.
Therefore, the second law of thermodynamics is also invalid for rotating Hayward black holes.
 
In order to check the third law of thermodynamics and weak cosmic censorship conjecture, we define a function $U$ and derive its form by using Eq.~(\ref{samphaybh}) and $r_e=M$,
\begin{equation}
    U\equiv \frac{\partial F(M,r_e)}{\partial M}+\frac{2}{M}=-\frac{2 M^3}{\left(M^2+2L^2\right)^2}+\frac{2}{M}.
\end{equation}
In terms of this definition, Eq.~\eqref{Tcon_5_1} becomes $U=0$.

\begin{figure}[htbp]
\centering
\includegraphics[width=0.6\linewidth]{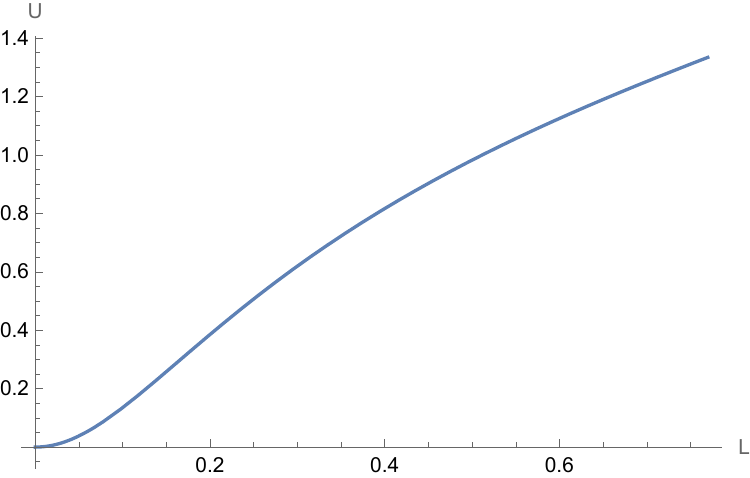}
\caption{The relationship between $U$ and $L$ in rotating Hayward black holes, where $M=1$ is set.}
\label{fig:Hayward_Third_Law}
\end{figure}

In Fig.~\ref{fig:Hayward_Third_Law}, we see that $U=0$ holds only for $L=0$, suggesting that rotating Hayward black holes do not satisfy the third law of thermodynamics and weak cosmic censorship conjecture, where $M=1$ is set without loss of generality.

\subsection{Double function case}
In this case, our discussions are based on the relationship between $H$ and $M$ in two categories, where $H$ does not depend on $M$ in the first category, and $H$ does depend on $M$ in the second category.

\subsubsection{Category of $H$ independent of $M$}
In this category, the necessary condition Eq.~\eqref{Fcon_1_4_4} of the first law of thermodynamics can be written as
\begin{equation}\label{Fcon_1_3}
    \frac{\partial H}{\partial r}\frac{\partial F}{\partial M}=-4,
\end{equation}
where $r\in[r_e,r_{\Hor(s)}]$, and the condition Eq.~\eqref{Fcon_2_4_4} remains unchanged.
For the second law of thermodynamics, the necessary conditions, Eqs.~\eqref{Scon_2_4_4} and \eqref{Scon_1_4_4}, become
\begin{equation}
    \frac{\partial H}{\partial r}\frac{\partial F}{\partial M}<0,
\end{equation}
and 
\begin{equation}
    H\frac{\partial F}{\partial r}\left(\frac{\partial H}{\partial r}\right)^{-1}-\frac{M}{2}\frac{\partial F}{\partial M}=1-F,
\end{equation}
where the range of $r$ is still $r\in[r_e,r_{\Hor(s)}]$.
For the third law of thermodynamics and weak cosmic censorship conjecture, the necessary condition is the same as Eq.~\eqref{Tcon_5_1}, but $r_e$ is taken to be zero if the ultimate state of black holes is a one-way wormhole.

Here we take the Kerr black-bounce solution as an example, whose static and spherically symmetric metric reads~\cite{Simpson:2018tsi}
\begin{equation}
ds^2=-\left(1-\frac{2M}{\sqrt{r^2+l^2}}\right)dt^2+\frac{dr^2}{1-\frac{2M}{\sqrt{r^2+l^2}}}+\left(r^2+l^2\right)\left(d\theta^2+\sin^2\theta d\phi^2\right),
\end{equation}
where $l$ is a positive parameter responsible for the regularization of the central singularity.
The black-bounce metric is interesting because it is a minimal one-parameter extension of the Schwarzschild metric.
This model converts the central singularity of Schwarzschild black holes into the throat of  wormholes after the introduction of parameter $l$, thereby achieving the purpose of connecting regular black holes and traversable wormholes. Recently, many regular black holes and traversable wormholes have been constructed~\cite{Junior:2023qaq,Pal:2022cxb,Lobo:2020ffi} by the black-bounce proposal.
However, the key issue is how to explain the black-bounce solutions physically, i.e., to find out the theory and matter that can yield such solutions. Now it is known~\cite{Bronnikov:2021uta} that this theory is the Einstein gravity coupled with matter, where the matter is the combination of a phantom scalar field and a nonlinear electrodynamics field. For the details~\cite{Bronnikov:2021uta},
the action reads,
\begin{equation}\label{eq:action_bb}
    I=\int{\sqrt{-g}\  d^4x\left(\mathcal{R}-2 g^{\mu\nu}\partial_\mu\phi_{\rm P}\partial_\nu\phi_{\rm P}-2V(\phi_{\rm P})-\mathcal{L}(\mathcal{F})\right)}
\end{equation}
where $\mathcal{L}(\mathcal{F})$ is the Lagrangian density of gauge-invariant nonlinear electrodynamics with the Faraday electromagnetic invariant, $\mathcal{F}\equiv F^{\mu\nu}F_{\mu\nu}$, and $\phi_{\rm P}$ is a phantom scalar field.
The Lagrangian density and the potential of a phantom scalar field take the following forms,
\begin{equation}\label{eq:L_bb}
    \mathcal{L}(\mathcal{F})=\frac{12Ml^2}{5\left(2q^2/\mathcal{F}\right)^{5/4}}
\end{equation}
and
\begin{equation}\label{eq:V_bb}
    V(\phi_{\rm P})=\frac{4M\cos^5\phi_{\rm P}}{5l^3},
\end{equation}
respectively, where $q$ is magnetic charge of free nonlinear electrodynamics.
And the Faraday electromagnetic invariant and the phantom scalar field read 
\begin{equation}\label{eq:F_bb}
	\mathcal{F}=\frac{2q^2}{(r^2+l^2)^2},
\end{equation}
and 
\begin{equation}\label{eq:phi_bb}
	\phi_{\rm P}(r)=\tan^{-1}\frac{r}{l}+{\rm const.},
\end{equation}
respectively.

The corresponding  metric with rotation is given~\cite{Mazza:2021rgq} by 
\begin{equation}
ds^2=-\left(1-\frac{2M\sqrt{r^2+l^2}}{\Sigma}\right)dt^2+\frac{\Sigma}{\Delta}dr^2-\frac{4Ma\sqrt{r^2+l^2}\sin^2\theta}{\Sigma}dtd\phi+\Sigma d\theta^2+\frac{A\sin^2\theta}{\Sigma}d\phi^2,
\end{equation}
where
\begin{subequations}
\begin{equation}
\Sigma=r^2+l^2+a^2\cos^2\theta,
\end{equation}
\begin{equation}
\Delta=r^2+l^2+a^2-2M\sqrt{r^2+l^2},
\end{equation}
\begin{equation}
A=(r^2+l^2+a^2)^2-\Delta a^2\sin^2\theta.
\end{equation}
\end{subequations}
Note that the above metric is consistent with Eq.~\eqref{eq:rotating_1} when $\Psi=\Sigma$ is chosen. According to $\Delta(r_{\rm H})=0$, we obtains the horizons, 
\begin{equation}
    r_{\rm H}=\sqrt{(M\pm\sqrt{M^2-a^2})^2-l^2}.
\end{equation}
When $l\in(0, M]$, the event horizon of an extreme black hole is located at $r_e=\sqrt{M^2-l^2}$, and the rotation parameter satisfies the relation: $a=M$.
When $l\in(M, 2M]$, the event horizon of the ultimate state is situated at $r_e=0$, and the rotation parameter satisfies the relation: $a=\sqrt{l(l-2M)}$, resulting in the degeneration of the Kerr black-bounce solution into a one-way wormhole.

Next, we consider the necessary conditions for the laws of thermodynamic and weak cosmic censorship conjectures.
Owing to 
\begin{equation}
     \frac{\partial H}{\partial r}\frac{\partial F}{\partial M}=-\frac{4r}{\sqrt{r^2+l^2}}\neq -4,
\end{equation}
in the range of $r\in[\sqrt{M^2-l^2}, \sqrt{4M^2-l^2}]$, see Eq.~\eqref{Fcon_1_4_4}, the first law of thermodynamics does not hold for the Kerr black-bounce solution. 

The necessary conditions for the second law take the forms, 
\begin{equation}
     \frac{\partial H}{\partial r}\frac{\partial F}{\partial M}=-\frac{4r}{\sqrt{r^2+l^2}}<0,\label{ncon2ndsec1}
\end{equation}
and 
\begin{equation}
    H\frac{\partial F}{\partial r}\left(\frac{\partial H}{\partial r}\right)^{-1}-\frac{M}{2}\frac{\partial F}{\partial M}=\frac{2M}{\sqrt{r^2+l^2}}=1-F,\label{ncon2ndsec2}
\end{equation}
see Eqs.~\eqref{Scon_2_4_4} and \eqref{Scon_1_4_4}, indicating that both Eq.~(\ref{ncon2ndsec1}) and Eq.~(\ref{ncon2ndsec2}) are valid simultaneously in the range of $r\in[\sqrt{M^2-l^2}, \sqrt{4M^2-l^2}]$.
Therefore, the second law of thermodynamics holds for the Kerr black-bounce solution.
This result shows that the area entropy can be used to define the Kerr black-bounce entropy, but the first law of thermodynamics still remains unsatisfied.

Finally, let us examine the third law and weak cosmic censorship conjecture for the Kerr black-bounce solution.
When $l\in(0, M]$, according to Eq.~\eqref{Tcon_5_1}, i.e.,
\begin{equation}
    \frac{\partial F(M,r_e)}{\partial M}=-\frac{2}{\sqrt{r^2+l^2}}=-\frac{2}{M},
\end{equation}
we obtain that the horizon of extreme configurations exists.
When $l\in(M, 2M]$, according to
Eq.~\eqref{Tcon_5_1} again, i.e., 
\begin{equation}
    \frac{\partial F(M,0)}{\partial M}=-\frac{2}{l}\neq-\frac{2}{M},
\end{equation}
we see that the horizon of ultimate states disappears.
However, the horizon undergoes a natural disappearance, i.e., the Kerr black-bounce solution changes to a one-way wormhole and then to a two-way wormhole.
As a result, the Kerr black-bounce solution satisfies the third law of thermodynamics and weak cosmic censorship conjecture.

\subsubsection{Category of $H$ dependent on $M$}
In this category, several necessary conditions of the laws of thermodynamics cannot be further simplified, so
our verification to  the laws depends on the specific forms of metrics.
Here we take rotating loop quantum gravity black holes as an example.

The metric functions of static seed black holes can be written~\cite{Brahma:2020eos} as
\begin{subequations}
    \begin{equation}
        F(r)=\left(1-\frac{2M}{\sqrt{r^2+4\lambda_k^{{2}/{3}}M^{{2}/{3}}}}\right)\frac{r^2+4\lambda_k^{{2}/{3}}M^{{2}/{3}}}{r^2+\lambda_k^{{2}/{3}}M^{{2}/{3}}},
    \end{equation}
    \begin{equation}
        H(r)=r^2+\lambda_k^{{2}/{3}}M^{{2}/{3}},
    \end{equation}
\end{subequations}
where the quantum parameter $\lambda_k$ originates~\cite{Bodendorfer:2019nvy,Bodendorfer:2019jay} from holonomy modifications.
This metric is static and spherically symmetric, which is obtained in terms of the effective equation of loop quantum gravity and is considered to be the quantum extension of Schwarzschild black holes.
In the rotating metric, the function $\Delta$ that determines horizons reads
\begin{equation}
    \Delta=FH+a^2=r^2+4\lambda_k^{{2}/{3}}M^{{2}/{3}}-2M\sqrt{r^2+4\lambda_k^{{2}/{3}}M^{{2}/{3}}}+a^2,
\end{equation}
and the corresponding horizons are located at 
\begin{equation}
    r_{\rm H}=\sqrt{\left(M\pm\sqrt{M^2-a^2}\right)^2-4\lambda_k^{{2}/{3}}M^{{2}/{3}}}.
\end{equation}
When $\lambda_k\in\left(0, M^2/8\right]$, the ultimate state of rotating loop quantum gravity black holes will stay at its extreme configuration with the horizon:  $r_e=\sqrt{M^2-4\lambda_k^{{2}/{3}}M^{{2}/{3}}}$, and the rotation parameter $a=M$.
When $\lambda_k\in\left(M^2/8, M^2\right]$, the ultimate state will be a one-way wormhole with $a=2M^{{1}/{3}}\sqrt{\lambda_k^{{1}/{3}}M^{{2}/{3}}-\lambda_k^{{2}/{3}}}$.

Next, we consider the necessary conditions for the laws of thermodynamics.
The first necessary condition, Eq.~\eqref{Fcon_1}, for the first law requires
\begin{equation}
    \frac{4 r \left[-5 \lambda_k ^{2/3} M+\lambda_k ^{2/3} \sqrt{4 (\lambda_k  M)^{2/3}+r^2}-M^{1/3} r^2\right]}{{M^{1/3}} \left[(\lambda_k  M)^{2/3}+r^2\right] \sqrt{4 (\lambda_k  M)^{2/3}+r^2}}=-4,
\end{equation}
but it is not satisfied within the range of $r\in[r_e, \sqrt{4M^2-4\lambda_k^{{2}/{3}}M^{{2}/{3}}}]$. Thus, the first law  is invalid for rotating loop quantum gravity black holes.

For the second law of thermodynamics, the necessary condition, Eq.~\eqref{Scon_1}, requires
\begin{equation}
    \frac{2 r (\lambda_k  M)^{2/3} \left[4 M-\sqrt{4 (\lambda_k  M)^{2/3}+r^2}\right]}{\left[(\lambda_k  M)^{2/3}+r^2\right] \sqrt{4 (\lambda_k  M)^{2/3}+r^2}}=0.
\end{equation}
It can only be satisfied when $\lambda_k=0$, in which case the rotating loop quantum gravity black holes turn back to Kerr black holes.
Therefore, the second law of thermodynamics is not satisfied by the rotating loop quantum gravity black holes.

At last, we examine the evolution of rotating loop quantum gravity black holes near an ultimate state.
When the ultimate state is an extreme configuration, the condition, Eq.~\eqref{Tcon_2}, changes into
\begin{equation}
    -\frac{6 \lambda_k ^{2/3}}{M^{1/3}}=0,
\end{equation}
which is unattainable under the circumstances, $M>0$ and $\lambda_k\neq0$.
Therefore, this case leads to disappearance of horizons when incident scalar particles are incoming, indicating a breakdown of both the third law and weak cosmic censorship conjecture.
If the final state is a one-way wormhole, the condition, Eq.~\eqref{Tcon_2}, changes into
\begin{equation}
    7\lambda_k^{1/3}-8M^{2/3}=0.
\end{equation}
It is only applicable to $\lambda_k=\left({8}/{7}\right)^3M^2$,  but $\lambda_k$ falls outside its reasonable range.
In this scenario, the rotating loop quantum gravity black holes may eventually transform into a two-way wormhole.
As a result, all the first, second, and third laws of thermodynamics are invalid in loop quantum gravity black holes.

\subsection{Summary for above applications}
To sum up, we list validity or invalidity  for three black hole models to obey the laws of thermodynamics and weak cosmic censorship conjecture in Tab.~\ref{Tab:1}.
The violation of the first law is evident in the three models. 
Meanwhile, both the Hayward and loop quantum gravity black holes fail to satisfy all the laws, indicating that rotating regular black holes may need new definitions of thermodynamic quantities and new laws of thermodynamics. 
For Kerr black-bounce solutions, while the first law is violated, both the second and third laws are still well fulfilled. 
This suggests that the area entropy possesses a certain degree of thermodynamic self-consistency. 
The primary challenge lies in modifying the first law of thermodynamics in order to achieve overall self-consistency for all the laws of thermodynamics.

\begin{table}[!ht]
	\centering
	\caption{This table shows whether the three specific models of rotating regular black holes satisfy the laws of thermodynamics and weak cosmic censorship conjecture, where  ``WCCC" is the abbreviation of  weak cosmic censorship conjecture.}
	\begin{tabular}{cccc}\hline
		
		& Hayward & Kerr black-bounce & Loop quantum gravity  \\ \hline
		
		The first law                           & no      & no                &    no  \\ \hline
		The second law                          & no      & yes               &    no  \\ \hline
		The third law                           & no      & yes               &    no  \\ \hline
		WCCC of extreme BHs   & no      & yes                  & no      \\ \hline
		WCCC of one-way wormholes   &       & no                 & no      \\ \hline
	\end{tabular}
	\label{Tab:1}
\end{table}

\subsection{Attempts to recover the laws of thermodynamics for rotating regular black holes}

The analyses made in the above three subsections show that rotating regular black holes break the laws of thermodynamics deduced from singular black holes. In fact, static regular black holes behave similarly, where
two modified approaches are mainly adopted: One is to modify~\cite{Nicolini:2008aj,Modesto:2008im,Kumar:2020uyz} the definition of entropy, and the other is to extend~\cite{Fan:2016hvf,Zhang:2016ilt,Rasheed:1997ns} phase spaces by treating regularized parameters as variables.
In the following we attempt to employ the two methods in the recovery of the laws of thermodynamics for rotating regular black holes.

\subsubsection{Modification of entropy}
Modifying the definition of entropy aims to establish the entropy that conforms to the first law of thermodynamics. 
As previously mentioned, if the first law of thermodynamics deduced from singular black holes holds for regular black holes, the second law also holds for regular black holes. 
Therefore, it can ensure the validity of the two laws to discover a suitable definition for entropy.
Let us assume that the entropy in question denoted by $S_M$ satisfies the first law of thermodynamics,
\begin{equation}
    T\dif S_M=\dif M-\Omega_\Hor\dif J,
\end{equation}
and it can be obtained through integration in the $(M, J)$ plane.
It is important to note that the entropy must be independent of the choice of integration paths.
Therefore, according to the path independence of curve integrals, the sufficient and necessary condition of a suitable entropy reads
\begin{equation}
    \frac{\partial}{\partial J}\left(\frac{1}{T}\right)=-\frac{\partial}{\partial M}\left(\frac{\Omega_\Hor}{T}\right).\label{suitentr}
\end{equation}
This condition is comparatively more flexible than the condition described by Eqs.~(\ref{Fcon_1_4_4}) and (\ref{Fcon_2_4_4}) because the latter requires not only a path-independent entropy but also an area entropy.

\begin{figure}[htbp]
	\centering
	\subfigure[Hayward]{
		\begin{minipage}[t]{0.4\linewidth}
			\centering
			\includegraphics[width=1\linewidth]{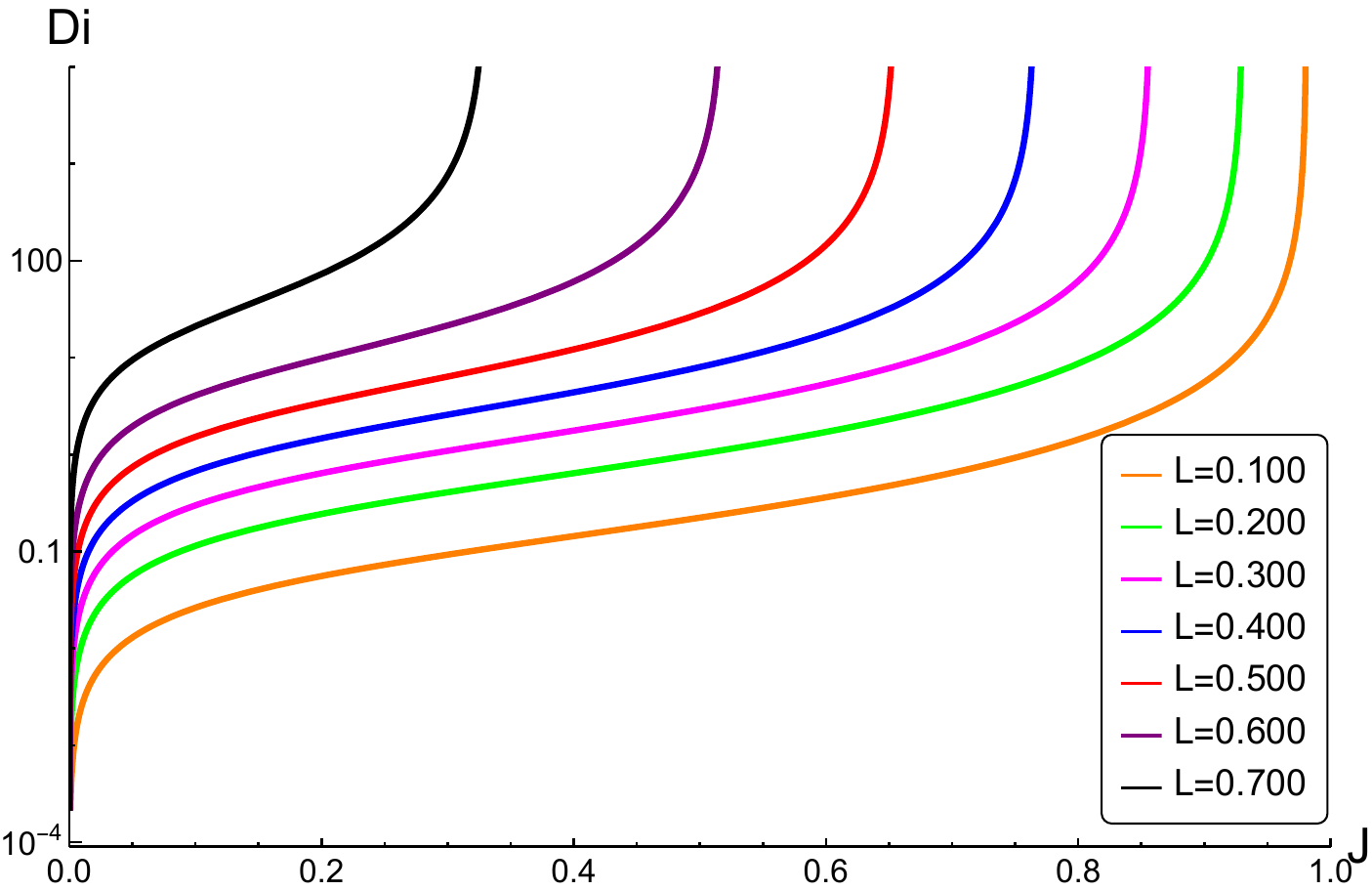}
		\end{minipage}
	}
	\subfigure[Kerr black-bounce]{
		\begin{minipage}[t]{0.4\linewidth}
			\centering
			\includegraphics[width=1\linewidth]{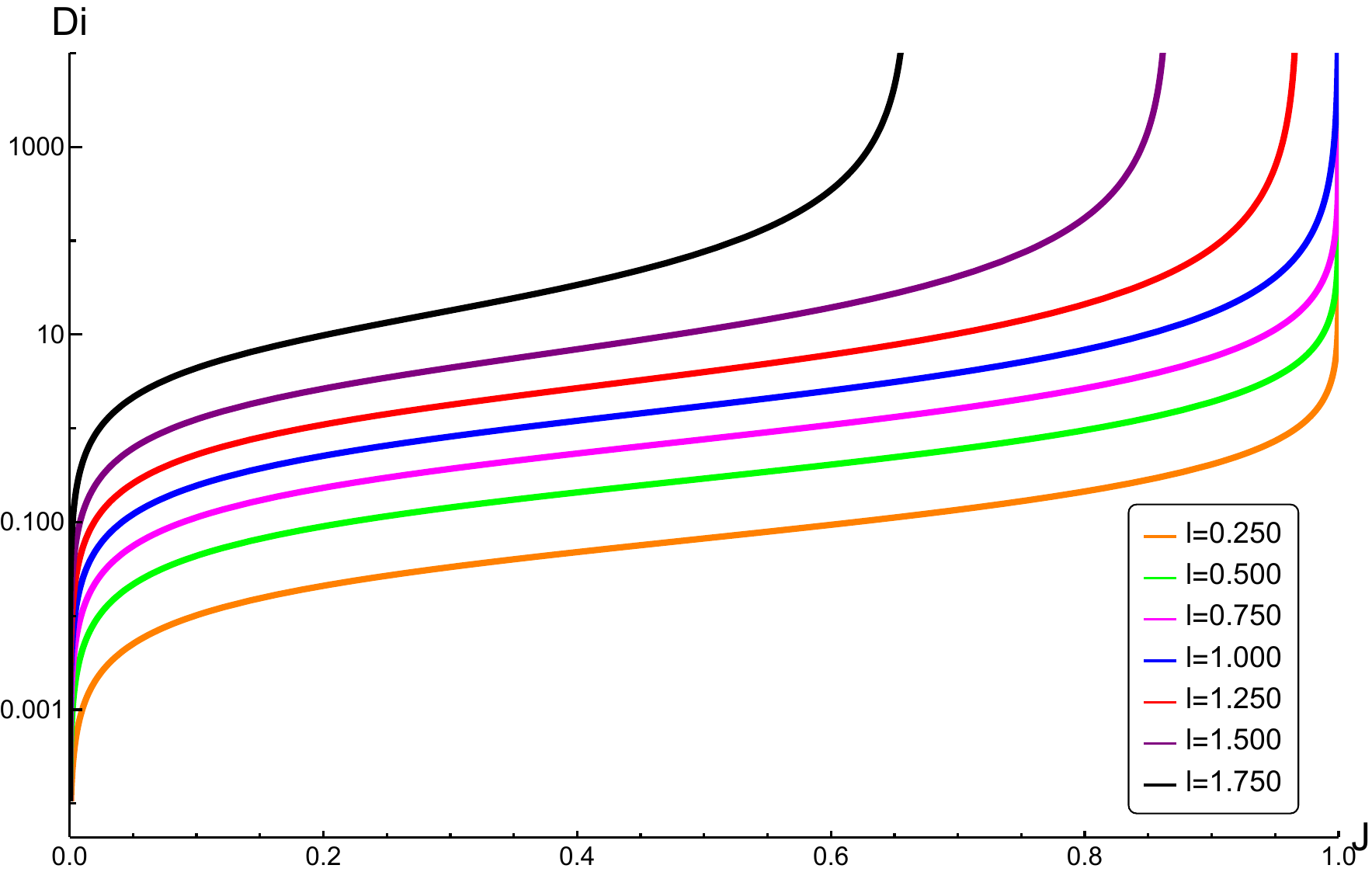}
		\end{minipage}
	}
	\vspace{-3mm}
	\subfigure[Loop quantum gravity]{
		\begin{minipage}[t]{0.4\linewidth}
			\centering
			\includegraphics[width=1\linewidth]{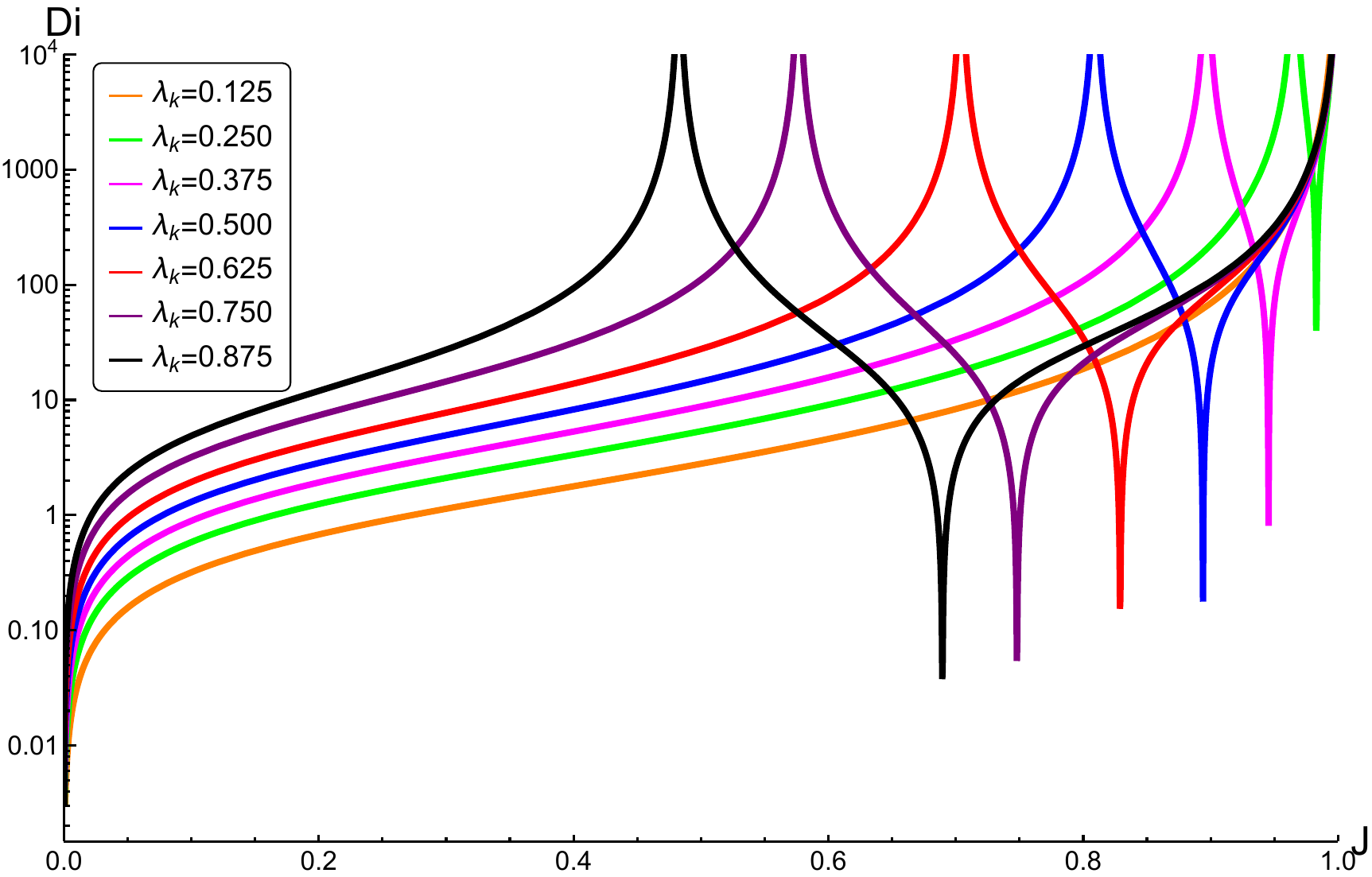}
		\end{minipage}
	}
	\caption{The relationship between $Di$ and the angular momentum under different regularization parameters in the three specific black hole models, where the black hole mass $M=1$ is set.}
	\label{fig:Di}
\end{figure}

Now we apply the  condition Eq.~(\ref{suitentr}) to the three specific models discussed above. 
By defining
\begin{equation}
    Di\equiv \left|\frac{\partial}{\partial J}\left(\frac{1}{T}\right)+\frac{\partial}{\partial M}\left(\frac{\Omega_\Hor}{T}\right)\right|,
\end{equation}
we draw diagrams shown in Fig.~\ref{fig:Di} in which $Di$ changes with respect to the angular momentum for the three modes with a given black hole mass $M=1$ but different regularization parameters.
Based on this figure, we conclude that no appropriate entropy exists for the fulfillment of the first law of thermodynamics because $Di$ is non-vanishing in the $(M,J)$ plane.
Hence, it is not feasible to require rotating regular black holes to satisfy the first law of thermodynamics solely by modifying entropy.

\subsubsection{Extension of phase spaces} 
We consider the regularization parameters, $(Y_1, Y_2, Y_3, \cdots, Y_n)$, as variables that vary during the process of particle incidence in order to extend a phase space.
According to the existence conditions of black hole event horizons before and after a particle incident process, we write the relationship among the first order derivatives of variables,
\begin{equation}\label{EPS_1}
    \dif r_\Hor=-\frac{\partial\Delta}{\partial M}\left(\frac{\partial\Delta}{\partial r}\right)^{-1}\Big|_{r=r_{\rm H}}\dif M-\frac{\partial\Delta}{\partial J}\left(\frac{\partial\Delta}{\partial r}\right)^{-1}\Big|_{r=r_{\rm H}}\dif J-\sum^{n}_{i=1}\frac{\partial\Delta}{\partial Y_i}\left(\frac{\partial\Delta}{\partial r}\right)^{-1}\Big|_{r=r_{\rm H}}\dif Y_i,
\end{equation}
and derive the corresponding change of entropy,
\begin{equation}\label{EPS_2}
    \dif S_\Hor=\frac{\partial S_\Hor}{\partial M}\dif M+\frac{\partial S_\Hor}{\partial J}\dif J+\frac{\partial S_\Hor}{\partial r_\Hor}\dif r_\Hor+\sum^{n}_{i=1}\frac{\partial S_\Hor}{\partial Y_i}\dif Y_i.
\end{equation}
By combining Eqs.~\eqref{EPS_1} and ~\eqref{EPS_2}, we obtain
\begin{equation}
    \dif S_\Hor=\tilde{A}\dif M+\tilde{B}\dif J+\sum^{n}_{i=1}\tilde{C}_i\dif Y_i,\label{entropydiff}
\end{equation}
where
\begin{subequations}
    \begin{equation}
        \tilde{A}=\frac{\partial S_\Hor}{\partial M}-\frac{\partial S_\Hor}{\partial r_H}\frac{\partial\Delta}{\partial M}\left(\frac{\partial\Delta}{\partial r}\right)^{-1},
    \end{equation}
    \begin{equation}
        \tilde{B}=\frac{\partial S_\Hor}{\partial J}-\frac{\partial S_\Hor}{\partial r_H}\frac{\partial\Delta}{\partial J}\left(\frac{\partial\Delta}{\partial r}\right)^{-1},
    \end{equation}
    \begin{equation}
        \tilde{C}_i=\frac{\partial S_\Hor}{\partial Y_i}-\frac{\partial S_\Hor}{\partial r_H}\frac{\partial\Delta}{\partial Y_i}\left(\frac{\partial\Delta}{\partial r}\right)^{-1}.
    \end{equation}
\end{subequations}
Here $\tilde{A}$ and $\tilde{B}$ play the same role as $\hat{A}$ and $\hat{B}$ in Eq.~\eqref{TFirst}, but $\tilde{C}_i$ are non-vanishing. Comparing Eq.~\eqref{TFirst} with Eq.~\eqref{entropydiff}, we deduce that the extension of phase spaces cannot play the role of revising the first law.
In conclusion, neither the modification of entropy nor extension of phase spaces can render the laws of thermodynamics for rotating regular black holes, indicating that 
further research is required to explore correct laws of thermodynamics for rotating regular black holes.

\section{Conclusion}\label{sec:conclusion}
In this paper, we establish the criteria for validity of thermodynamic laws and weak cosmic censorship conjecture in rotating regular black holes by examining the process of neutral scalar particles' incidence into a rotating regular black hole. 
In this process, we calculate mass, charge, and other conserved quantities of black holes in general, and give the formula of Hawking temperature for rotating regular black holes. 
By examining the relationships among these quantities, we evaluate whether rotating regular black holes satisfy the laws of thermodynamics deduced from singular black holes or not. Moreover,
we find that a complementary relationship between the third law of thermodynamics and weak cosmic censorship conjecture exists only when the ultimate state of a rotating regular black hole is an extreme configuration. 
Alternatively, the ultimate state of a rotating regular black hole is a one-way wormhole, such that
the rotating regular black hole evolves into a two-way wormhole without violating the third law but leading to disappearance of event horizons, i.e., leading to invalidity of the weak cosmic censorship conjecture.
In order to give a deeper understanding of such an evolution, we need to establish the thermodynamics of wormholes in a way that connects the thermodynamic states of rotating regular black holes before evolution to those states after evolution.


In addition, for three specific models of rotating regular black holes we verify their compliance with the thermodynamic laws and weak cosmic censorship conjecture.
However, as shown in Tab.~\ref{Tab:1}, two of the three models do not satisfy all the laws of thermodynamics, suggesting that a reestablishment of thermodynamics is necessary for them.
Fortunately, the Kerr black-bounce solution satisfies the second and third laws of thermodynamics. 
In the attempt to recover the laws of thermodynamics for rotating regular black holes, we resort the redefinition of entropy and extension of phase spaces. 
However, these two methods do not work, indicating the unusual property of rotating regular black holes.
Such an unusual property implies the necessity of employing alternative approaches that are distinct from those employed in static and spherically symmetric regular black holes, in order to recover the thermodynamic laws applicable to rotating regular black holes.
Meanwhile, if a black hole is treated as a thermodynamic system, the self-consistency of the first, second and third laws must fully be guaranteed.
	The significance of our results is that one may establish the  self-consistent thermodynamic laws
	by taking into account the inconsistencies we have revealed.

Finally, we propose some ideas that are helpful for a deeper understanding of rotating regular black holes.
\begin{itemize}
    \item To redefine the conserved quantities of a rotating regular black hole. 
    The aforementioned conserved quantities, such as mass and angular momentum, are defined in an entirety of spacetime. 
    If we just consider the region inside a horizon of a rotating regular black hole as a thermodynamic system, we may constrain the scope of integration within a horizon, which
    would affect the configurations of conserved quantities and thus the corresponding thermodynamic system.
   \item To give a new algorithm for the construction of a rotating regular black hole. 
   As a mathematical technique, the NJA transforms a static and spherically symmetric black hole into a rotating and axially symmetric one. 
   Owing to its nonphysical defects, such as a complex radial coordinate, we may require a more suitable algorithm for constructing a rotating regular black hole.
   \item To consider the reaction caused by particle incidence. 
   Our present  discussions  neglect the spacetime reaction during particle incidence. 
   If we consider the impact of particle incidence on field equations, the original relationships among conserved quantities may be altered, leading to modifications in thermodynamic laws.
\item 
To introduce the quantum correction in black hole models. 
It has been shown~\cite{Reuter:2010xb} that the first law of thermodynamics for some rotating black holes can be self-consistent by limiting the running Newton coupling under the asymptotically safe gravity. 
This method provides an alternative way to recover the first law of thermodynamics for the rotating regular black holes we have analyzed. It is an interesting issue to extend the method to the recovery of the second and third laws. 
\end{itemize}

\paragraph{Acknowledgments}
This work was supported in part by the National Natural Science Foundation of China under Grant No. 12175108.




\bibliographystyle{unsrturl}
\bibliography{references}

\end{document}